\documentstyle[12pt,epsfig]{article}

\begin{document}

\bibliographystyle{unsrt} 

\leftline{FNT/T-97/10}

\begin{center}
{\Large\bf Six-Fermion Calculation of \\}
{\Large\bf Intermediate-mass Higgs Boson Production  \\}
\vskip  6pt
{\Large\bf at Future $e^+ e^-$ Colliders}
\end{center}

\noindent
\begin{center}
Guido MONTAGNA$^{a,b}$, Mauro MORETTI$^{c}$, \\
Oreste NICROSINI$^{b,a}$  and Fulvio PICCININI$^{b,a}$ 
\end{center}

\noindent
\begin{center}
$^a$ Dipartimento di Fisica Nucleare e Teorica, 
Universit\`a di Pavia, Italy \\
$^b$ INFN, Sezione di Pavia,  Italy  \\
$^{c}$ Dipartimento di Fisica, Universit\`a di Ferrara and INFN, 
Sezione di Ferrara, Italy \\
\end{center}

\begin{abstract}
{\small The production of an intermediate-mass Higgs boson in 
 processes of the kind $e^+ e^- \to 6$~fermions at the energies of future linear 
colliders is studied. 
The recently developed and fully automatic algorithm/code {\tt ALPHA} is used
to compute the tree-level scattering amplitudes for the reactions 
$e^+ e^- \to \mu^+ \mu^- \tau^- \bar\nu_{\tau} u \bar d, \mu^+ \mu^- 
e^- \bar\nu_{e} u \bar d$. The code has been interfaced with the 
Monte Carlo program {\tt HIGGSPV/WWGENPV}, properly adapted to 6-fermion production, 
in order to provide  realistic results, both in the form of cross sections and event samples 
at the partonic level. Phenomenological results, that incorporate 
the effects of initial-state radiation and beamstrahlung, are shown and 
commented, emphasizing the potentials of full six-fermion calculations 
for precise background evaluation as well as for detailed studies of the 
fundamental properties of the Higgs particle. }
\end{abstract}

\vskip 24pt \noindent
E-mail: \\
montagna@pv.infn.it \\
moretti@axpfe1.fe.infn.it \\
nicrosini@pv.infn.it, nicrosini@vxcern.cern.ch \\
piccinini@pv.infn.it \\

\vfil
\leftline{FNT/T-97/10}

\leftline{May 16, 1997}
\vfill\eject

\normalsize
\section{Introduction}

The investigation of the mechanism of electroweak symmetry breaking through the 
search for  the Higgs particle constitutes one the main tasks of the experiments
at present and future colliders. Present-day efforts at LEP2 are expected   
to reach the mass limit $m_h \approx \sqrt{s} - 100$~GeV and 
therefore will be 
 unable to pursue the search for the Higgs boson in the interesting range of 
intermediate mass, i.e. $M_Z \leq m_h \leq 2 M_Z$. This mass interval is actually 
 particularly attractive since a number of theoretical arguments~\cite{thlim}, 
 as well as 
fits to precision electroweak data~\cite{blon}, provide indication for a weakly-coupled 
Higgs boson lighter than, say, 200~GeV. On the other hand, the discovery 
of the Higgs boson in the intermediate-mass interval might pose 
severe problems at 
the future large hadron collider LHC, in spite of continuous progress in 
detector performances and research strategies~\cite{atlascms} as well as in 
updating theoretical predictions~\cite{kms}. 
 
From this point of view, the next-generation of $e^+ e^-$ linear colliders (NLC) 
could be of great help in the complementary search for a Higgs boson with 
intermediate mass, as well as in the determination 
of its fundamental properties and couplings to SM particles.
At the NLC, for ``moderate'' centre of mass (c.m.) energies up to $0.5$~TeV,
the main production mechanism for the Higgs boson
is the Higgs-strahlung 
process $e^+ e^- \to Z H$.
When considering the $M_H>135\div140$~GeV range,
the above Higgs-strahlung 
process gives rise to a six-fermion final state, originating from 
$Z$ decaying into $f \bar f$ pairs and 
 Higgs boson  decaying predominantly into $WW$ pairs,  
with the subsequent decay of each $W$ into leptonic $l \nu_l$ or
hadronic $u \bar d$ final states. Therefore, the full calculation of the
rate of intermediate-mass Higgs boson production at the NLC 
faces with the problem of a full calculation of $2 \to 6$ scattering 
amplitudes, including Higgs signal and background processes in the electroweak
theory. The same difficulty applies to other very important research streams 
of the NLC, namely $t {\bar t}$ and three-vector bosons production, that 
both generate six-fermion final states.

A few   calculations 
of such processes have been very recently performed, and a short account of the 
existing approaches can be found  in~\cite{orcr}. Although different 
in several computational and numerical aspects, all these strategies rely upon 
the calculation of the very large number of Feynman diagrams involved (of the order of hundreds), 
exploiting in particular the experience accumulated during the last few years
in the computations of the processes $e^+ e^- \to 4$~fermions developed for 
$W-$pair and ``light'' Higgs boson production at LEP2~\cite{wwwg,smwg,hwg,wweg,dpeg}.       
Concerning these recent calculations 
of specific  processes of the kind 
$e^+ e^- \to 6$~fermions  ($6f$), in~\cite{to} the full set 
of diagrams for the semi-leptonic 
processes $e^+ e^- \to e^- \bar\nu_{e} 4q$ and $e^+ e^- \to \mu^- \bar\nu_{\mu} 4q$ 
have been computed and physical distributions of interest for 
{\it top}-quark and $WWZ$ physics have been analyzed in detail. Similar calculations are in progress 
following a fully computerized approach for the automatic computation 
of the Feynman diagrams by means of the {\tt GRACE} system~\cite{kek}. 
Previous calculations of
helicity amplitudes relative to multi-particle production processes and relevant  
for collider phenomenology can also be found in the  literature~\cite{hz}.

Here  an alternative attack strategy to the
 problem is presented,  using a theoretical algorithm (and the corresponding code) 
recently proposed in the literature and known as {\tt ALPHA}~\cite{alpha}. 
For a given Lagrangian, this algorithm, of iterative nature, allows to compute numerically, 
in a fully automatic way, the tree-level scattering amplitudes, 
without using the Feynman diagrams, and it turns out to be 
particularly powerful for the calculation of those processes involving 
a high number of final-state particles. 
The resulting  code has already been  applied with success to the calculation of the rates of 
multi-particle production reactions of interest 
for the LEP2 and NLC experimental programme, such as $e^+ e^- \to 4$~fermions, 
$e^+ e^- \to 4$~fermions+$\gamma$~\cite{alpha,cmg} and 
$\gamma \gamma \to \bar{\nu_e} e^- u \bar d$~\cite{ggm}. In particular, 
in the case of four-fermion production processes the numerical results 
obtained with {\tt ALPHA} have been compared in detail with those 
of independent formulations based on standard computational techniques, 
showing excellent agreement and thus providing a stringent test of the 
algorithm and of the resulting code~\cite{smwg,wweg}. In order to produce realistic results, 
the algorithm has been interfaced with the 
Monte Carlo program {\tt HIGGSPV/WWGENPV}~\cite{higgspv,wwgenpv}, 
developed for $WW$ and Higgs boson physics in  processes with four fermions in the final state in  the
context of LEP2 physics~\cite{lep2}, and properly adapted to 
$6f$ production.  The computational  tool developed can work 
both as an integrator of weighted events, providing cross sections for any given experimental set up, 
and as a generator of unweighted events, providing event samples suited for physics analysis and/or  
detector simulation. 

To the best of the authors'  knowledge, a full six-fermion 
calculation of intermediate-mass  higgs boson production in $e^+ e^-$ 
collisions and relative phenomenological analysis have not yet 
been performed.  In this paper  the results obtained 
for two different channels are presented, trying to emphasize the 
usefulness of a complete $6f$ calculation from the point of view of 
precise background evaluations in Higgs boson searches as well as for the 
determination of the fundamental properties (mass, spin, etc.) of this particle.
Recent reviews on the main aspects of Higgs  boson phenomenology at future 
$e^+ e^-$ linear colliders can be found in~\cite{r0,r1,r2,rd,rz,rmp}. Moreover, also 
full calculations for processes of the kind $e^+ e^- \to bbWW$, $bbZZ$ and $4jets + W$, 
of interest for  Higgs boson searches at NLC, 
but with on shell final-state vector bosons, can be found in~\cite{smo,bmm}. 
  
The paper is organized as follows. In Section 2,  the 
details of the calculation are presented, 
considering intermediate-mass Higgs boson production in the 
six-fermion reactions $e^+ e^- \to \mu^+ \mu^- \tau^- \bar\nu_{\tau} u \bar d, 
\mu^+ \mu^- e^- \bar\nu_{e} u \bar d$; a sample of illustrative numerical results
is shown and commented in Section 3 and the main conclusions as well as 
possible perspectives 
 are drawn in Section 4.

\vskip 12pt\noindent
\section{Six-fermion Calculation}

In order to study
the intermediate-mass Higgs boson production at the NLC 
in channels  that  are 
as free as possible from large background contamination,  
only   the processes 
\begin{eqnarray}
e^+ e^- \to \mu^+ \mu^- \tau^- \bar\nu_{\tau} u \bar d, 
\mu^+ \mu^- e^- \bar\nu_{e} u \bar d
\end{eqnarray} 
are considered, 
which differ for their content of 
electroweak backgrounds but are both completely 
unaffected by QCD backgrounds. 

In the following, some details concerning the strategies 
followed for the calculation of the physical amplitudes and the phase 
space of the above $2 \to 6f$ processes will be described.

As said in the introduction, 
to  compute  the matrix elements, 
 a recently proposed technique
({\tt ALPHA}) has been employed. It doesn't require the evaluation of Feynman graphs 
and allows to obtain automatically, according to an iterative procedure, 
the tree-level scattering amplitudes of any given lagrangian. This is 
achieved by exploiting the relation between the generator of 
one-particle irreducible Green functions and the generator of the connected 
Green functions. 
The interested reader is referred to  the
original literature~\cite{alpha} for 
a detailed description of the method.
Here it is worth  pointing out that 
\begin{itemize}

\item  the procedure, which is entirely automatic, has been 
checked successfully against the precise predictions existing 
for the processes $e^+ e^- \to 4f$~\cite{smwg,wweg} and applied 
to obtain original results for the rate of the reactions 
$e^+ e^- \to 4f + \gamma$~\cite{cmg} and 
$\gamma \gamma \to 4$~fermions~\cite{ggm};

\item to calculate the amplitude for a certain process, an input file is 
provided, where one has simply to specify the type of the process and the 
total number of particles involved in intial and final states.
For the pure QED case the required input is an integer vector
$N_{prt}\equiv(n_1,\dots,n_6)$ where $n_1,\dots,n_6$
are the numbers of initial $e^-$, final $e^+$, final $e^-$, initial $e^+$,
initial and final photons respectively; once this vector is initialized and 
the proper kinematics is provided {\tt ALPHA} returns the scattering
matrix elements;\footnote{For the standard model 
the only difference is that the vector $N_{prt}$
has much more entries (a pair for each SM particle). }

\item the present  calculation is the first application of {\tt ALPHA} to the 
computation of $e^+ e^- \to 6f$ processes; this enables, in particular, 
to check yet untested parts of the input lagrangian (i.e.~the coupling of 
the higgs boson to the $W$ and $Z$ bosons and some of the 
quartic non-abelian gauge boson couplings) and to show the potentials 
of the algorithm for multi-particle production processes involving a 
very large number of diagrams. 

\end{itemize}  

\begin{figure}[hbtp]
\begin{center}
\epsfig{file=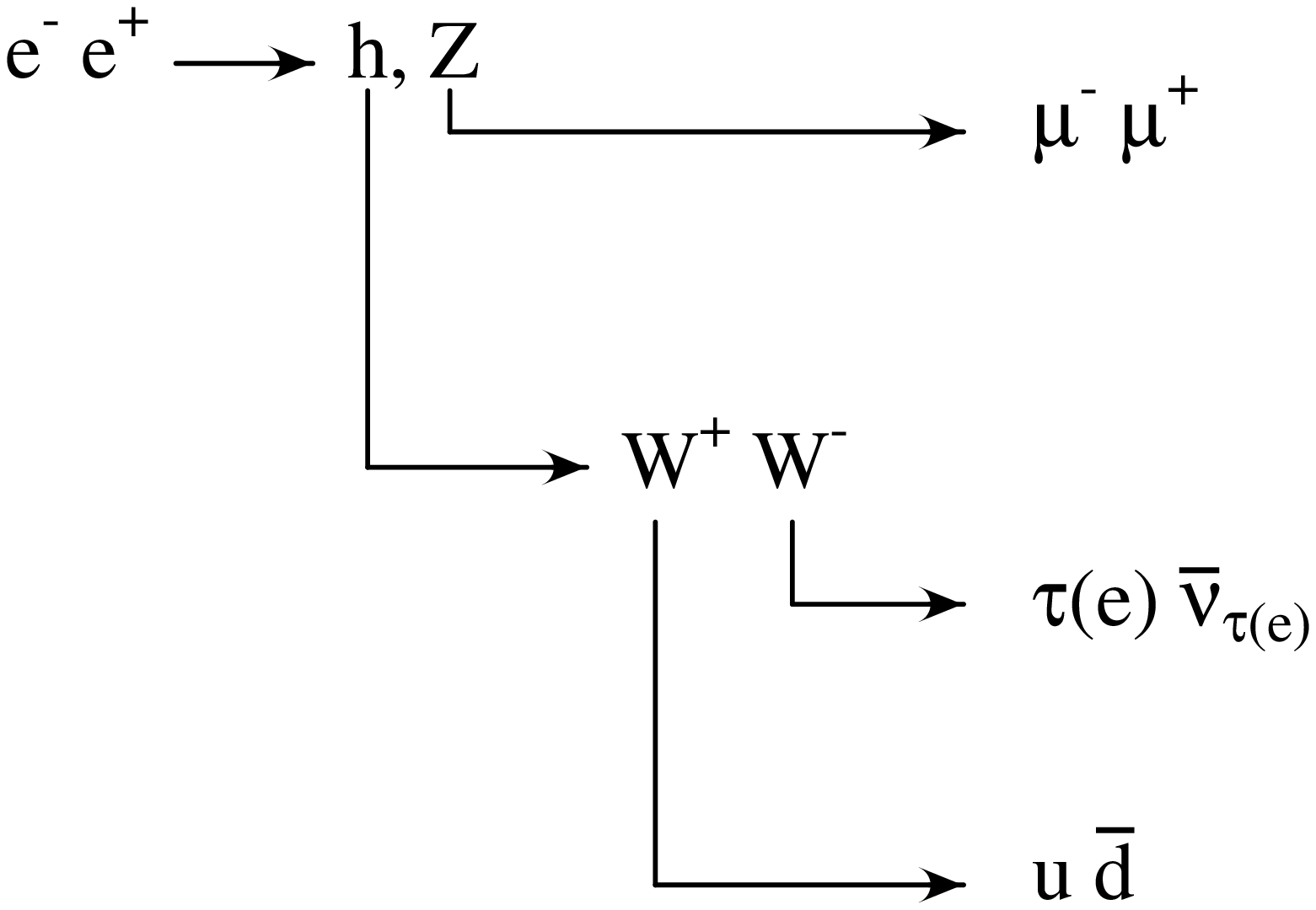,height=6truecm, width=9truecm}
\end{center}
\caption{The  decay chain in  the channel  $e^+e^-  \to \mu^+
\mu^- u \bar d \tau^-(e^-) \bar \nu_{\tau(e)}$. }
\label{fig:scheme}
\end{figure}

The kinematics of the $2 \to 6f$ processes 
has been treated generating the 6-body phase space recursively.
This choice is particularly useful since  processes 
where a particle decays into other particles which subsequently decay  
are being considered (see fig.~\ref{fig:scheme}).
In particular,  denoting by $p_{-(+)}$ the
four-momentum of the incoming electron (positron), $q_i$, $i = 1,\ldots , 6$, the final-state
outgoing four-momenta and $P = p_- + p_+$, the phase space can be written as
\begin{eqnarray}
&& d \Phi_6 (P; q_1, \ldots, q_6) = (2 \pi)^{12} d \Phi_2 (P; Q_Z, Q_h) 
d \Phi_2 (Q_Z; q_1,q_2) \nonumber \\
&& d \Phi_2 (Q_h; Q_{W^+}, Q_{W^-}) d \Phi_2 (Q_{W^+}; q_3, q_4) 
d \Phi_2 (Q_{W^-}; q_5, q_6) \nonumber  \\
&& d Q_Z^2 d Q_h^2 d Q_{W^+}^2 d Q_{W^-}^2 , 
\end{eqnarray}
where $Q_Z  = q_1 + q_2$, $Q_h = \sum_{i=3}^6  q_i$, $Q_{W^+} = q_3 + q_4$ and $Q_{W^-} = q_5
+ q_6$.
Globally, fourteen independent variables (including a trivial 
overall azimuthal angle) are required; they have been chosen according to the 
following scheme: 
\begin{itemize}
\item 
four invariant masses 
$Q_Z^2$, $Q_h^2$, $Q_{W^+}^2$, $Q_{W^-}^2$
\item
 five $\vartheta$- and $\varphi$-angle pairs, 
in the rest frame of each decaying ``particle'', namely in the
 frames where $\vec{P} = 0$, ${\vec{Q}_Z}=0$,
${\vec{Q}_h}=0$,
${\vec{Q}_{W^+}}=0$
and ${\vec{Q}_{W^-}}=0$, respectively; once these variables are generated, 
the four-momenta of the outgoing fermions are derived in each ``rest" frame and 
then boosted back to the laboratory frame. 
\end{itemize}
The choice of the kinematics has been motivated by the dynamics of the Higgs boson signal; the  same phase
space is also conveniently employed for the background processes. 

In order to obtain reliable phenomenological results for the 
energies of the future $e^+ e^-$ colliders, the lowest-order 
calculation of the matrix  elements has to be supplemented with the potentially 
large effects of energy losses due to the emission of initial state photons, caused by  
the mutual interaction of the colliding electrons and positrons 
at the ``parton" level (bremsstrahlung or ISR), as well as to the synchrotron 
radiation generated by the strong electromagnetic interaction 
between dense bunches of particles (beamstrahlung). Both the contribution of
ISR and beamstrahlung have been incorporated in the physics simulation in order 
to provide as realistic as possible predictions. The task has been accomplished by interfacing 
{\tt ALPHA} with the  Monte Carlo program {\tt HIGGSPV/WWGENPV}, properly adapted to treat 
$6f$ production processes.  This allowed to exploit as much as possible the expertise developed 
in the context of the LEP2 workshop~\cite{lep2}, concerning 4-fermion production processes. 
In particular,   the method of QED structure functions  (SF)~\cite{sf} is used to account for ISR 
according to the factorized formula
\begin{eqnarray}
\sigma (s) = \sum_i \int_0^1 d x_1 \, d x_2 \, D(x_1,s) D(x_2,s) 
d [PS] {{d \sigma} \over {d [PS]}} {{w_i} \over {W}} , 
\end{eqnarray}
where $D(x,s)$ is the electron SF, $d [PS]$ denotes the volume 
element of the 6-body  phase space, $d \sigma / d [PS]$ is the tree-level 
hard scattering density in the c.m. frame, built by means of the total scattering
amplitude as returned by {\tt ALPHA} and of the proper jacobian factors; $w_i$ and  
$W = \sum_i w_i$ are weights introduced to increase the efficiency of the 
Monte Carlo integration, namely by incorporating the weights $w_i$ in the integration measure 
according to a  multi-channel   importance sampling~\cite{james80}; the random number generator 
employed is {\tt  RANLUX}~\cite{ranlux}.  
Beamstrahlung corrections are simulated using the parameterizations of 
such effects recently implemented in the library {\tt CIRCE}~\cite{circe}, namely according to the
formula 
\begin{equation}
\sigma (s) = \sum_i  \int_0^1 d z_1 d z_2 D_{BS} (z_1,z_2; \sqrt{s}) 
 \int_0^1 d x_1 \, d x_2 \, D(x_1,s) D(x_2,s) 
d [PS] {{d \sigma} \over {d [PS]}} {{w_i} \over {W}} , 
\end{equation}
where $D_{BS}$ is the beamstrahlung distribution.\footnote{\footnotesize The {\tt TESLA} accelerator code 
has been used. The library {\tt CIRCE} provides the beamstrahlung parameterization for some  fixed 
$\sqrt s$, in particular $\sqrt s = 350$ and 500~GeV. 
For one of the c.m. energies considered in the following simulations, namely $\sqrt s =
360$~GeV, the parameterization  relative to 350~GeV has been used, as explicitly allowed in the original
documentation. } 
When ISR and/or beamstrahlung are  taken into account, the four-momenta of  the final-state
particles are  additionally  boosted  to the laboratory frame, in order to take into account the
c.m. boost  generated by such effects. At present, ISR is treated in the strictly collinear
approximation.  

\begin{figure}[hbtp]
\begin{center}
\epsfig{file=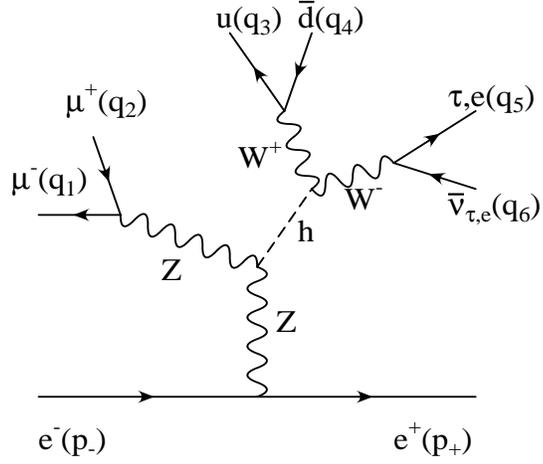,height=6truecm, width=7truecm}
\end{center}
\caption{The Feynman diagram for the Higgs boson signal in the channel $e^+e^-  \to \mu^+
\mu^- u \bar d \tau^-(e^-) \bar \nu_{\tau(e)}$. }
\label{fig:higgs_signal}
\end{figure}

\begin{figure}[hbtp]
\begin{center}
\hbox{
\hbox{\epsfig{file=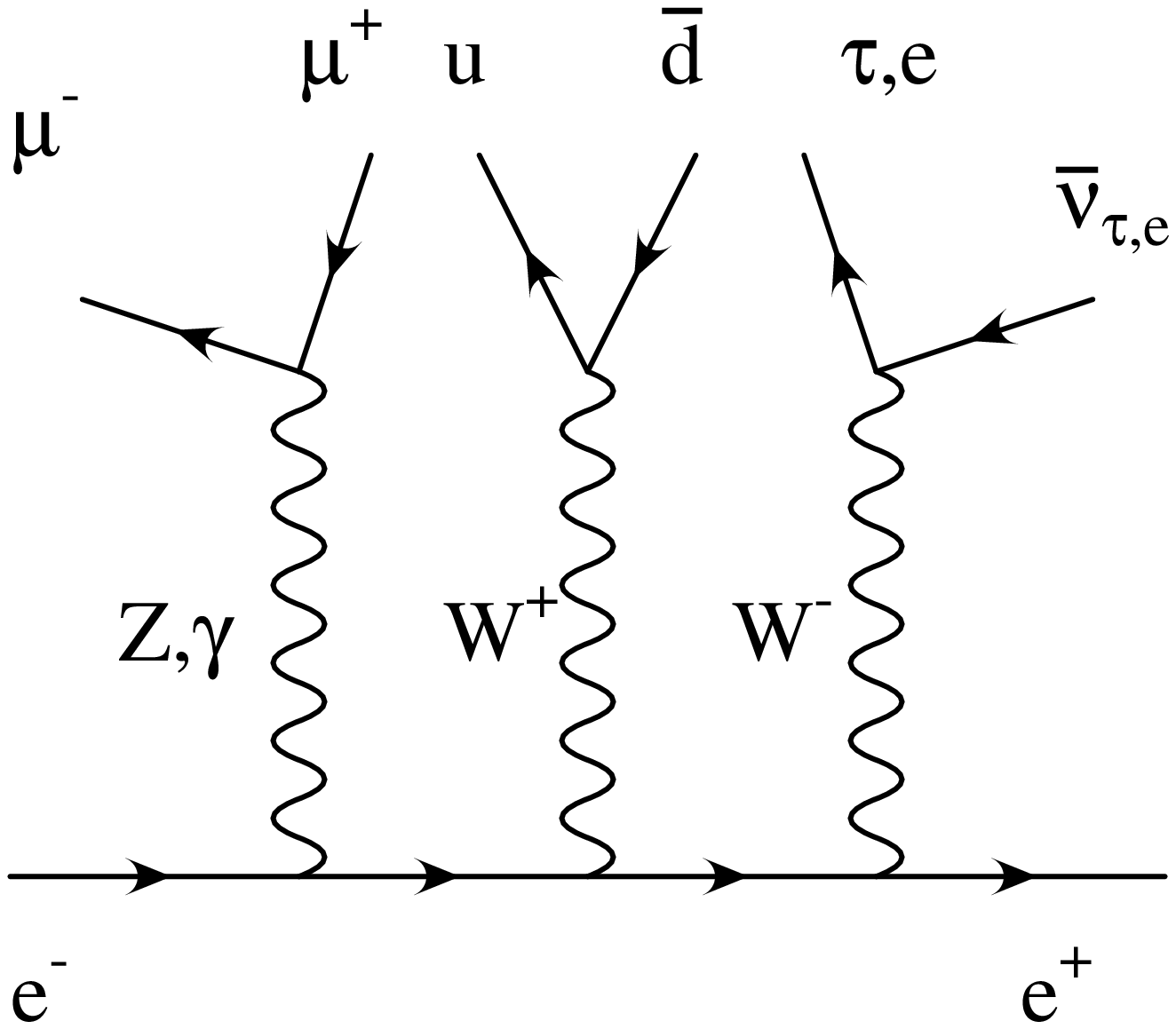,height=4truecm, width=5truecm}}
\hbox{\hskip 3.3truecm}
\hbox{\epsfig{file=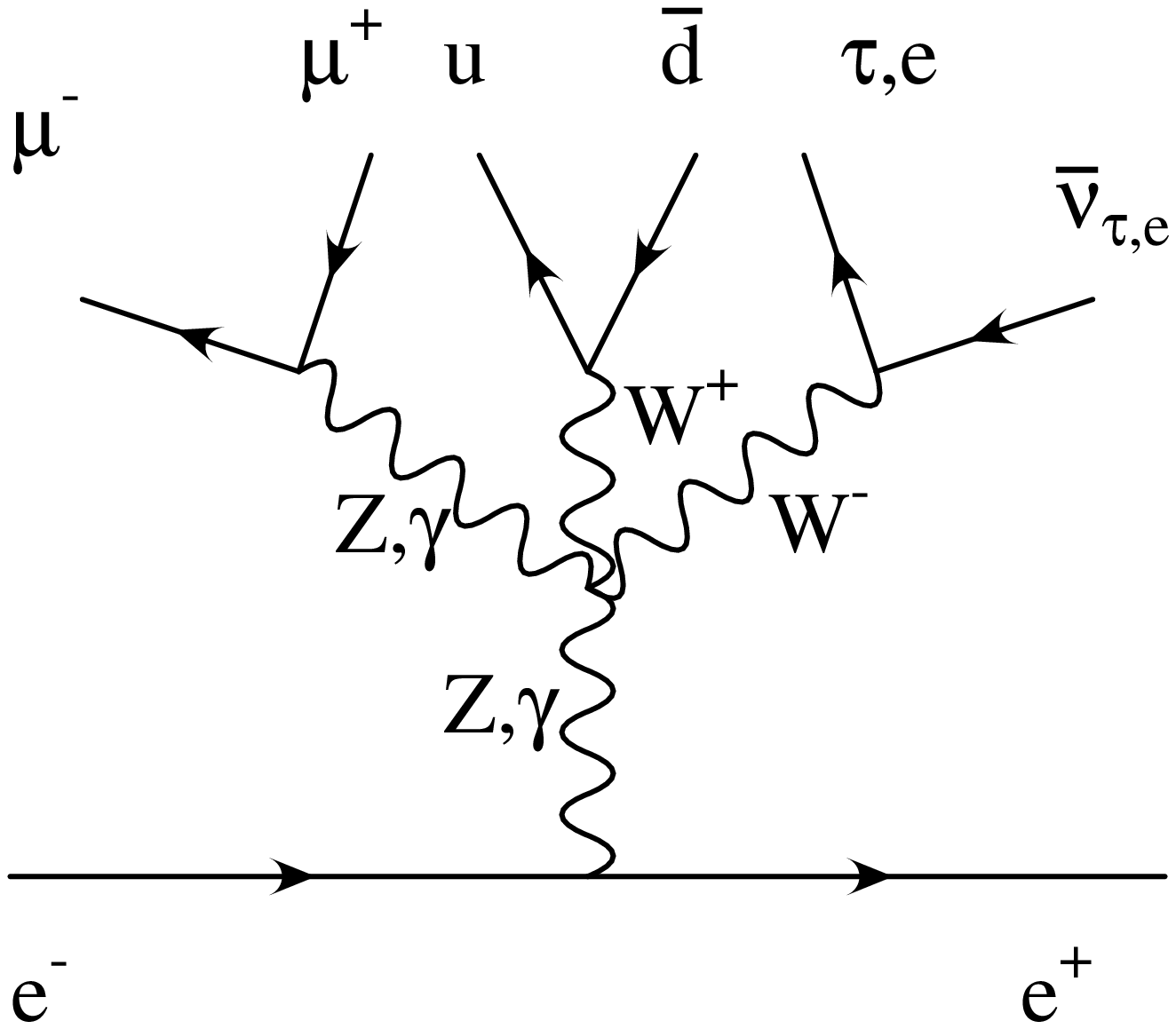,height=4truecm, width=5truecm}}
}
\end{center}
\caption{Feynman diagrams for typical background processes  in the channel $e^+e^-  \to \mu^+
\mu^- u \bar d \tau^-(e^-) \bar \nu_{\tau(e)}$. }
\label{fig:cbckg}
\end{figure}

\begin{figure}[hbtp]
\begin{center}
\hbox{
\hbox{\epsfig{file=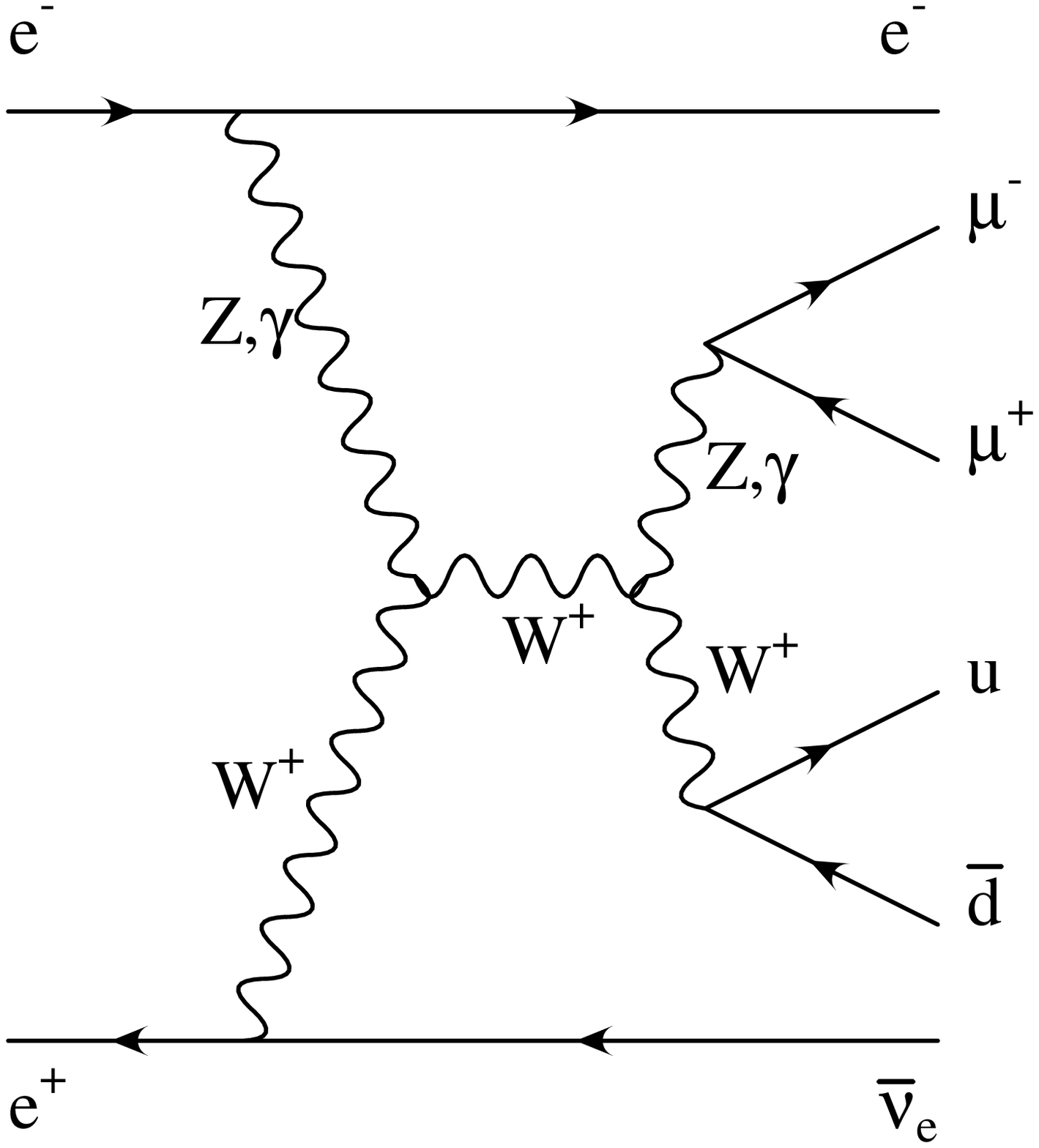,height=4truecm, width=5truecm}}
\hbox{\hskip 3.3truecm}
\hbox{\epsfig{file=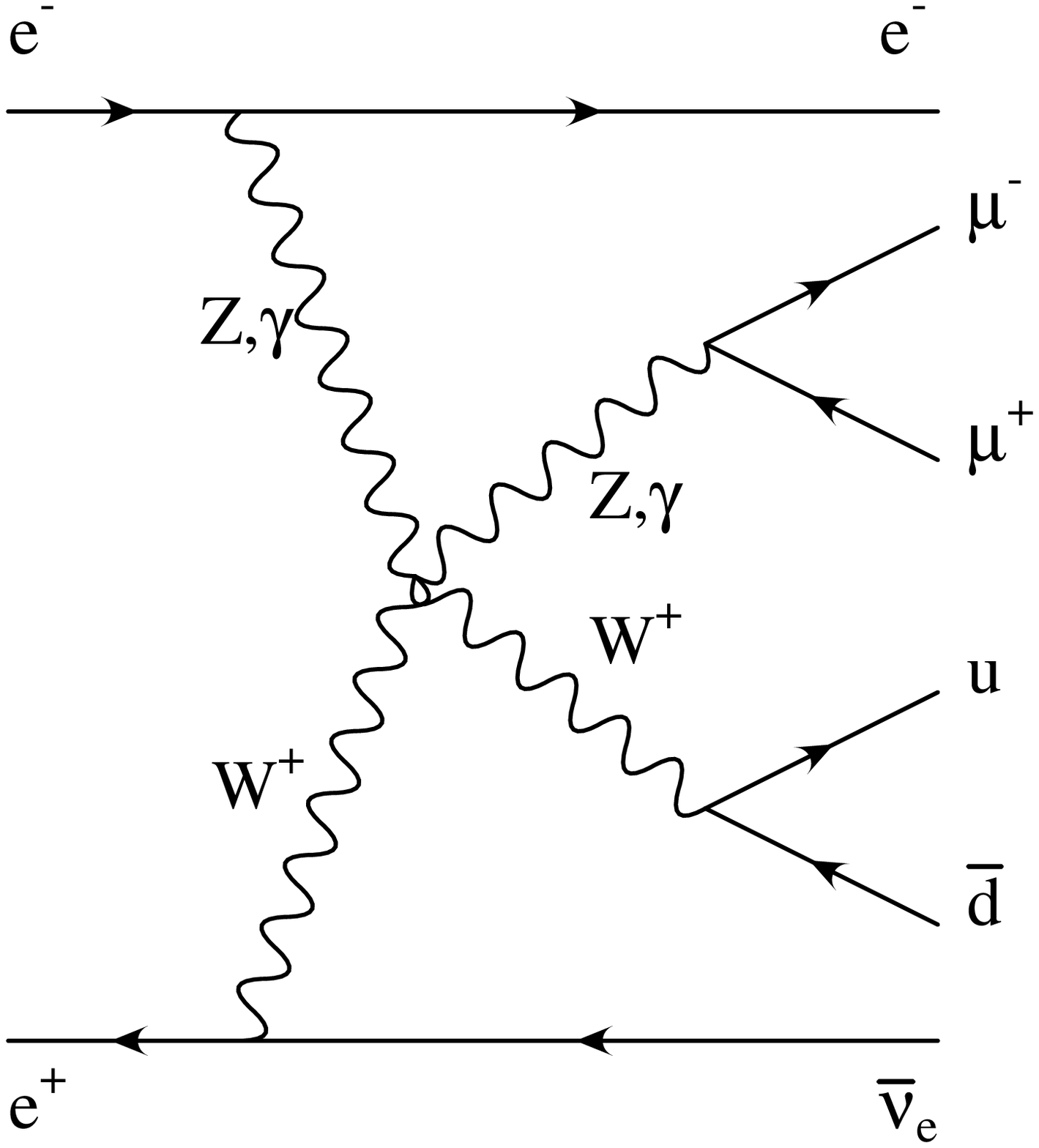,height=4truecm, width=5truecm}}
}
\end{center}
\caption{Additional Feynman diagrams for typical background processes  
in the channel $e^+e^-  \to \mu^+
\mu^- u \bar d e \bar \nu_{e}$. }
\label{fig:ebckg}
\end{figure}

A technical but important item in the numerical  calculation  is  the choice of the weight
functions  $w_i$ employed for the implementation  of the importance sampling.  This  choice has to
be driven by the physics involved. When considering only the Higgs boson signal (see
fig.~\ref{fig:higgs_signal}), it  has to  be
observed that muon pairs come   from the $Z$  boson decay, while the remaining four final-state 
fermions come  from the Higgs boson decay. Moreover, the hadronic pair comes  from the $W^+$ boson
decay and the additional  leptonic  pair from the $W^-$ boson decay. 
This means that  the most populated phase space  region is in the surroundings of $Q^2_Z \simeq
M^2_Z$, $Q^2_h \simeq m^2_h$ and $Q^2_{W^\pm} \simeq  M^2_W$. 
When considering the leading background 
processes common to both the channels  examined  (see for instance fig.~\ref{fig:cbckg}), 
the muon pairs can come also, for instance, from photon conversion; 
moreover, $Q^2_h$ is no more peaked around
$m^2_h$;  the hadronic and the additional leptonic pairs still come from $W$'s decay. At last, when
considering background processes typical of the channel with  the electron in the final state  (see
for instance fig.~\ref{fig:ebckg}) the $e \nu$ leptonic pair invariant mass is  no more peaked 
around $M^2_W$. Following  these considerations, the importance sampling  adopted  has the following
features: 
\begin{itemize} 
\item  $Q^2_h $  is sampled piecewise, according to a Breit-Wigner (BW)  
density around $Q^2_h  = m^2_h$
and a flat density  elsewhere; the relative  weights of  the BW and flat densities  are
tuned  according  to the signal/background ratio for  any  given $\sqrt{s}$; 

\item  $Q^2_Z$ is sampled piecewise,  according to a BW  density around $Q^2_Z =  M^2_Z$, 
according to $1 / Q^2_Z$ in the lower  tail and  a flat density in  the upper one; 
 
\item $Q^2_{W^\pm}$ are always  sampled  according to a  BW  density; 

\item  since a  realistic event selection requires  a  minimum scattering angle for  the leptons, no
particular sampling  has  been performed in the final-electron forward region; in particular, a 
minimum angle  of $5^\circ$ between the final electron and  the  incoming beams  has  been imposed,
in such a way that the $t$-channel ``singularity'' (see
for instance fig.~\ref{fig:ebckg})  is  excluded.   
\end{itemize}

The ``soft-photon singularity'' of  the  electron  SF's is sampled in the standard way, as can 
be found  in the literature~\cite{higgspv,wwgenpv}.  

CPU performances have not been the  main concern of the study. Anyway, it is  worth saying that 
the Monte Carlo code developed,  in the worst case,  generates around 3500 unweighted
events in 8h CPU  and, as integrator of weighted events, provides a cross section with
a statistical error of around 1\% in 3h CPU on a DEC ALPHA  station. Such performances are sufficient for
any realistic simulation.

\vskip 12pt\noindent
\section{Numerical results and discussion} 

In order to  produce the numerical results shown in the present section, the input  parameters
used are the Fermi coupling constant $G_F$, the $Z$ boson mass $M_Z =  91.1888$~GeV and the $W$
boson mass $M_W  = 80.23$~GeV. The weak  mixing  angle and the QED coupling constant are then 
computed as $\sin^2 \vartheta_w = 1 - M^2_W / M^2_Z$ and $4 \pi \alpha = 
g^2 \sin^2 \vartheta_w $, where $g^2  = 8 M_W^2  G_F / \sqrt{2}$. Given these input parameters,
all the other relevant quantities, such as the $Z$ and $W$ bosons widths, have been computed at
the tree-level approximation
and the massive boson propagators 
are chosen as $\sim 1/(p^2 - M^2 + i\Gamma M)$.  
This choice of input parameters has been performed
in order to  avoid problems concerning violations of the Ward identities,
or, more precisely, to confine all possible gauge violation
effects into the inclusion of the finite fixed widths 
for the massive bosons. 

As far as the Higgs boson width is concerned, in the present simulation the following 
contributions have been taken  into account: the fermionic contributions $h \to \mu\mu,
\tau\tau, cc, bb$, where the hadronic widths have been QCD corrected taking into account the
running quark mass effects~\cite{hwg}; the gluonic contribution $h \to g g$ according
to~\cite{hwg}; the vector boson contribution $h \to V^* V^*$, according to ref.~\cite{kniel}. 
 
The result has been tested to be $U(1)_{em}$ gauge-invariant, by comparing the results obtained
with two different inverse photon propagators in the effective lagrangian implemented in 
{\tt ALPHA}. {\it A priori}, one could expect $U(1)$ gauge invariance problems to be severe 
in particular in the channel with the electron in the final state. Anyway, 
such potential problems~\cite{bhf1} are avoided  requiring a realistic event selection, as described 
above. Concerning the $SU(2)$ gauge invariance, it has been  pointed  out in~\cite{bhf2} 
that the fixed-width scheme, used in the present study, is not $SU(2)$ gauge invariant and could in 
principle be responsible of a bad  high-energy behaviour in processes involving six fermions in  the 
final state. The numerical relevance of the $SU(2)$ gauge violations  of the (fixed widths)  results 
has then been checked by 
comparing them with results obtained in the so called ``fudge scheme''~\cite{fudge}, that is by 
construction gauge-invariant, and finding them compatible at the per cent level. 

\begin{figure}[hbtp]
\begin{center}
\hbox{
\hbox{\epsfig{file=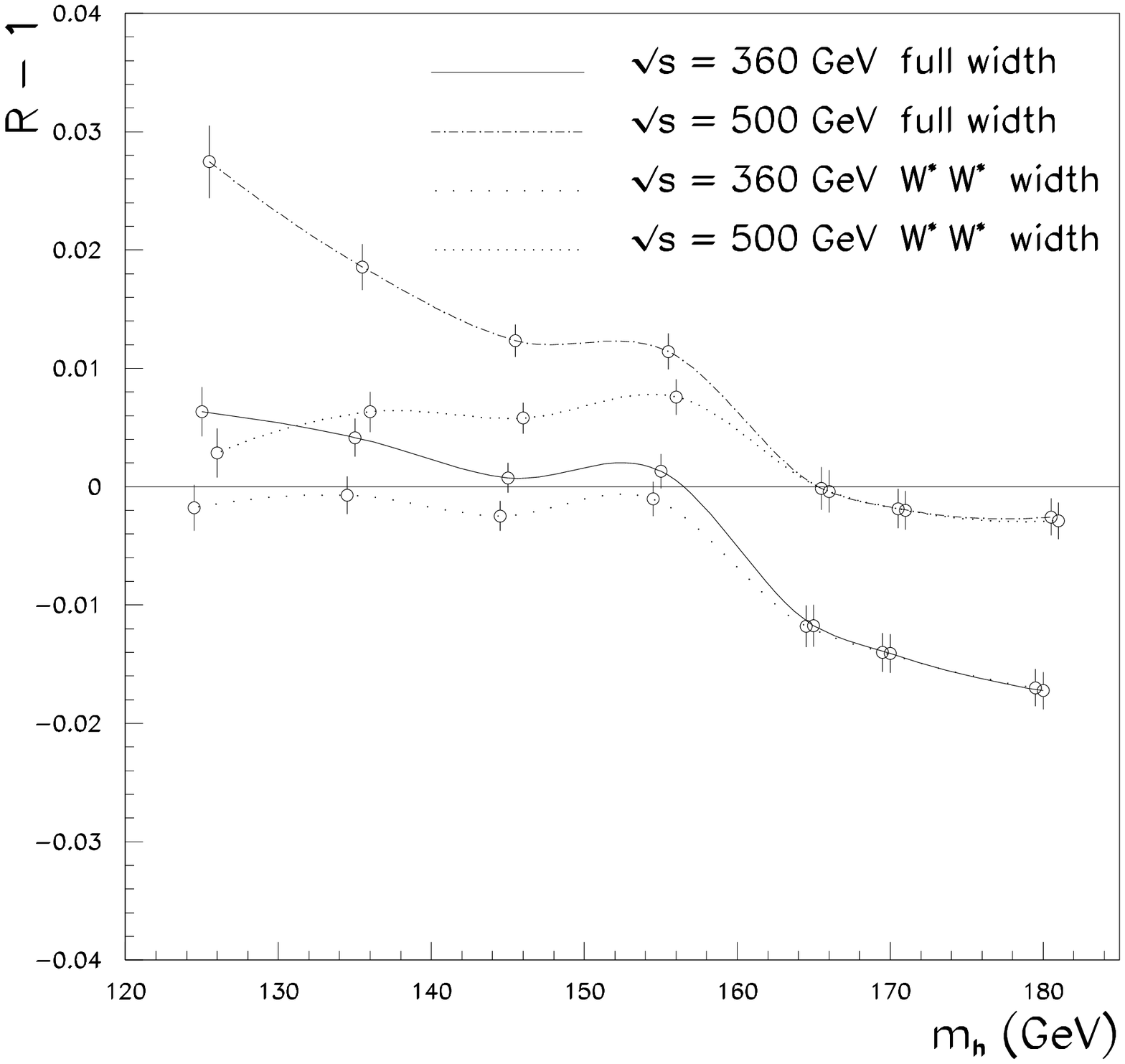,height=6.3truecm, width=6.3truecm}}
\hbox{\hskip 1truecm}
\hbox{\epsfig{file=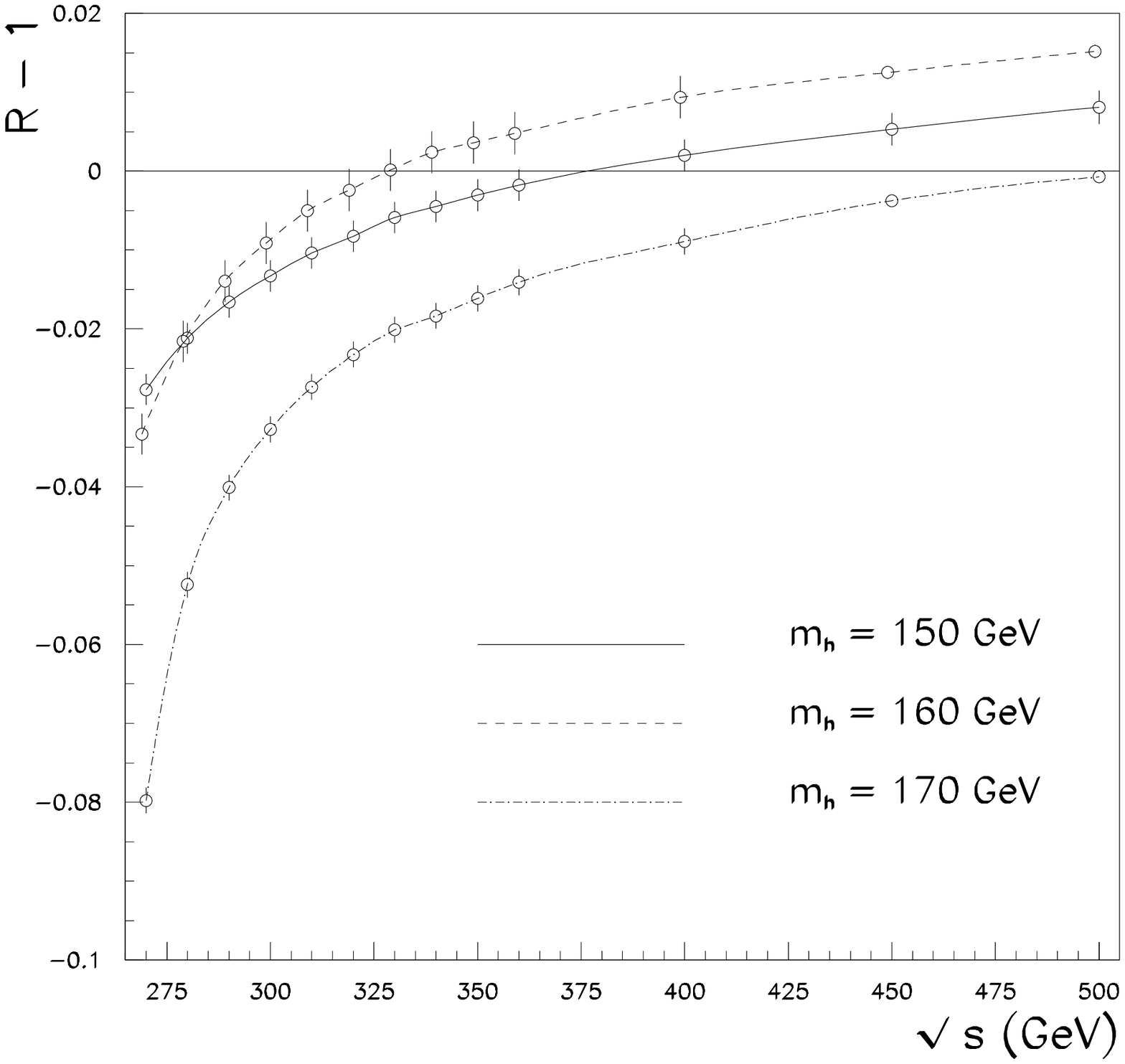,height=6.3truecm, width=6.3truecm}}
}
\end{center}
\caption{The comparison between the results of the full six fermion simulation and the results
obtained in the narrow-width approximation,  as function of the Higgs  boson mass $m_h$ 
and of  the c.m.  energy. }
\label{fig:comparison}
\end{figure}

All the numerical results, but those shown in fig.~\ref{fig:comparison}, have been obtained by
imposing the following {\it  a priori} cuts: the invariant mass of the muon pair larger than
20~GeV; the invariant mass of the hadronic system larger than 10~GeV; the angle between the
charged leptons and the beams larger than $5^\circ$.  

\begin{figure}[hbtp]
\begin{center}
\hbox{
\hbox{\epsfig{file=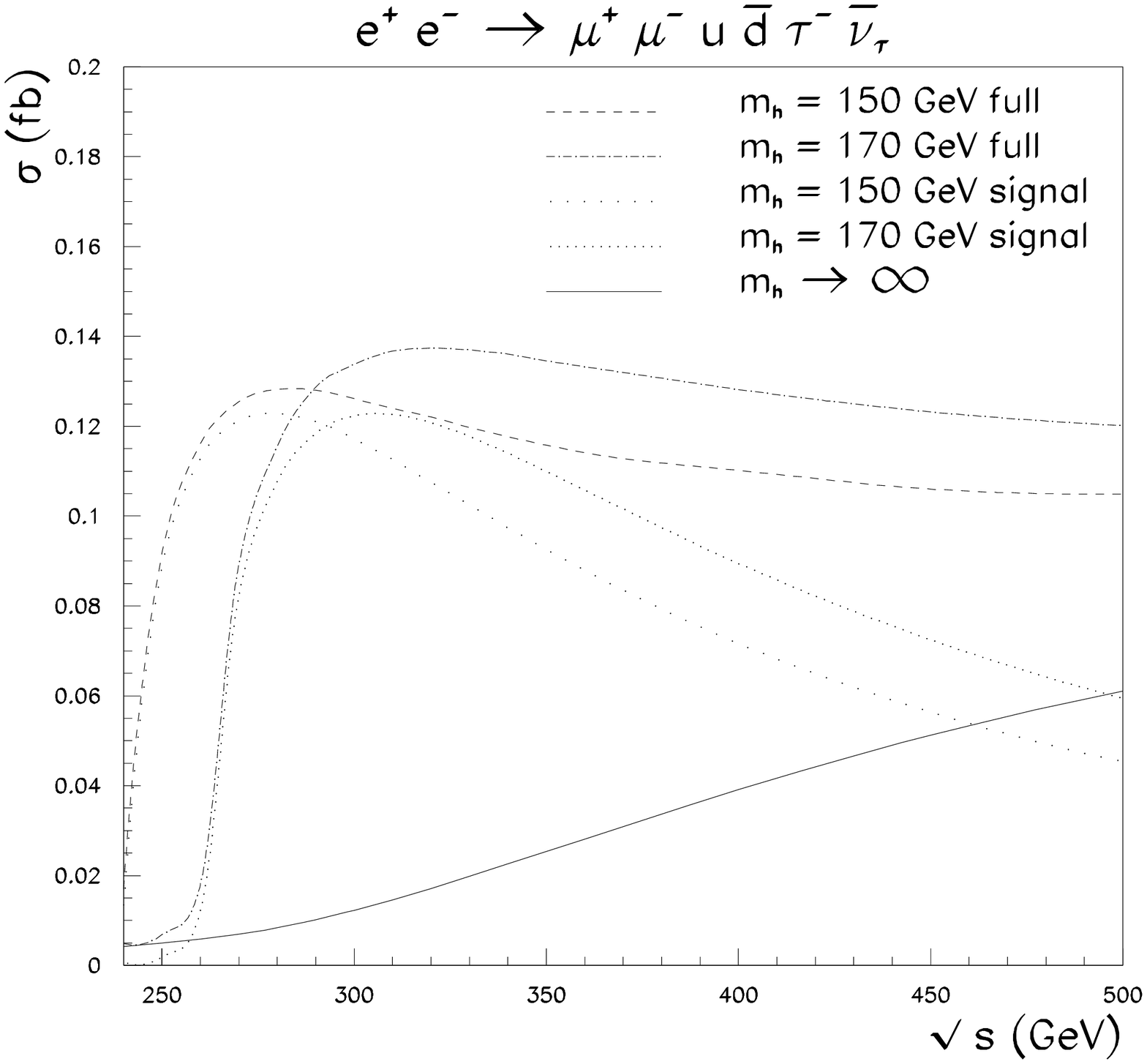,height=6.3truecm, width=6.3truecm}}
\hbox{\hskip 1truecm}
\hbox{\epsfig{file=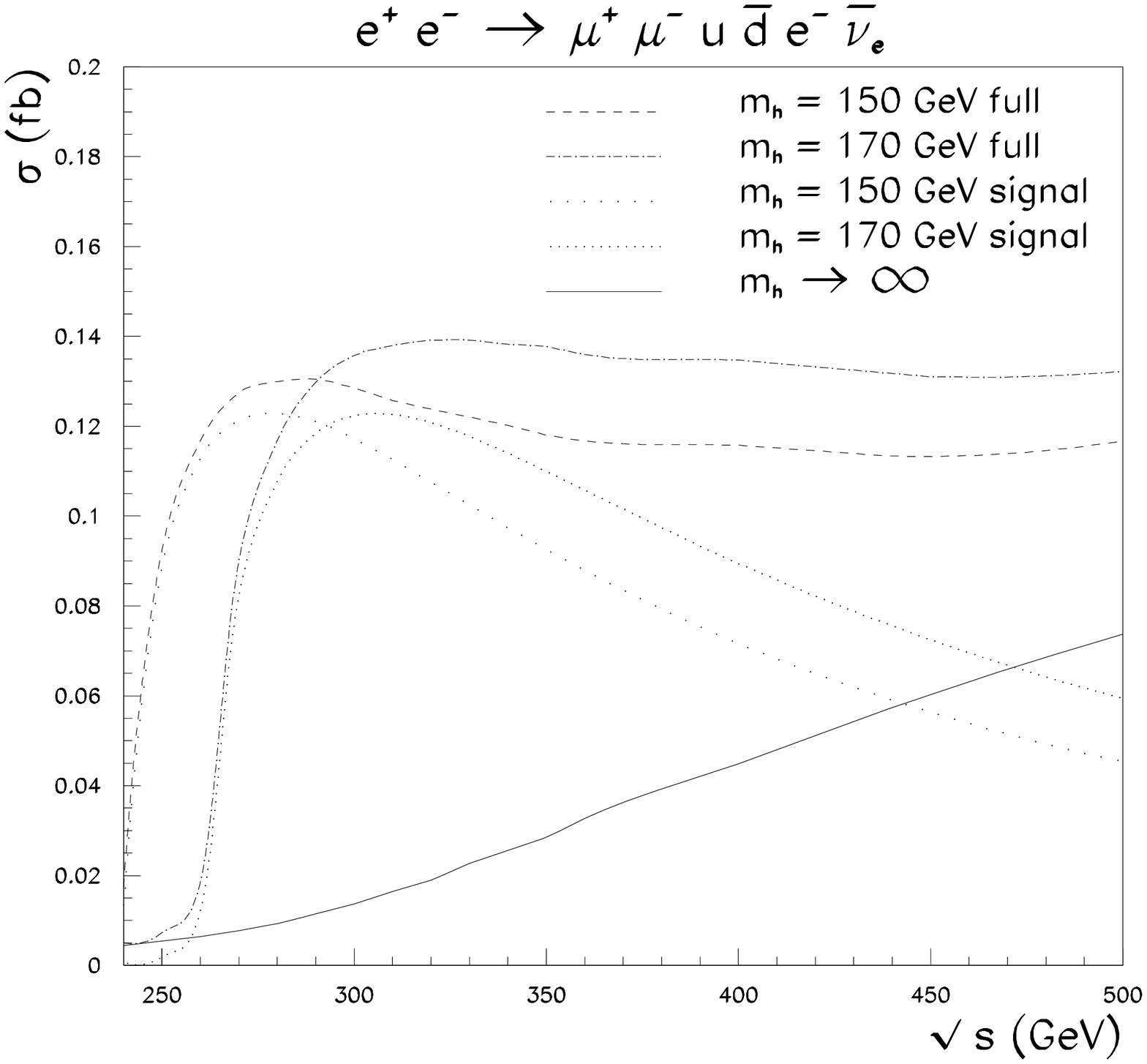,height=6.3truecm, width=6.3truecm}}
}
\end{center}
\caption{The cross sections for the channels $e^+ e^- \to \mu^+ \mu^- 
\tau^- \bar\nu_{\tau} u \bar d$
and $e^+ e^- \to \mu^+ \mu^- e^- \bar\nu_{e} u \bar d$ as functions of the c.m.  energy. }
\label{fig:totxsect}
\end{figure}

The results shown in figs.~\ref{fig:comparison} and \ref{fig:totxsect} 
have been obtained by using the  program as a Monte Carlo integrator of weighted events.

Figure~\ref{fig:comparison} shows the comparison between the fully extrapolated 
cross sections for the Higgs
boson signal alone, as obtained by means of the present complete  $6f$ calculation on the one hand, 
and within 
the narrow-width approximation (NWA) according to ref.~\cite{zerwas93} on the other one. 
The two  windows show the
comparison as a function of the Higgs boson mass, for two fixed $\sqrt{s} = 360$ and 500~GeV, 
and as a function of the c.m. energy, for three fixed Higgs boson masses $m_h =
150$, 160 and 170~GeV, respectively. In both  cases the relative deviation $R = \sigma_{6f} /
\sigma_{NWA} - 1$ is considered. In  particular, in the first plot two cases are  considered,
namely the case in which the total Higgs boson decay width has  been used in the full $6f$
calculation (solid and dashed lines), and the case in which the Higgs boson width in the full
$6f$ calculation is identified  with $\Gamma (h \to W^*W^*)$ (dotted and sparsely dotted
lines). The dotted and sparsely  dotted lines approach  zero in the region of ``small'' Higgs
boson masses: this represents a check that the  $6f$ calculation reproduces the results
already present in the literature for  the NWA. The solid and dashed lines do not approach
zero, because of the presence of the fermionic and gluonic  contributions to the total Higgs
boson width,  but  this is the realistic case. Anyway, the off-shellness effects are contained
within a few per cent, as can  be also seen in the second plot. 

Figure~\ref{fig:totxsect} shows the comparison, at the 
tree-level approximation, between the full cross section, the separate 
contribution of the Higgs boson signal and the  background
only, for two values of the Higgs boson mass $m_h = 150, 170$~GeV and as
a function of the c.m. energy. The first  plot shows the results for  the channel 
$e^+ e^- \to \mu^+ \mu^- \tau^- \bar\nu_{\tau} u \bar d$, the second one for the channel 
$e^+ e^- \to \mu^+ \mu^- e^- \bar\nu_{e} u \bar d$. As a first comment, it has to be noticed that, for
the event selection considered, the channel containing the electron in the final state is not
very much different from the one containing the $\tau$ lepton; this in spite of the fact  that
the electron channel receives contributions also from $t$-channel processes of the kind shown in
fig.~\ref{fig:ebckg}. The $t$-channel background processes start to become visible only in the
high-energy tail of the plot, say around $\sqrt{s} \simeq 500$~GeV. 
The statistical error due to the
numerical integration is of the order of 0.5\% and hence invisible for the scale adopted in the plots. 
In  both channels, the full calculation is
consistent with the incoherent sum of signal and background, at the statistical level considered. 
Signal/background interference effects can become relevant for resolutions better than  1\%. 
It has to be noticed that, in both cases, 
$\sqrt{s} \simeq 500$~GeV represents the turning point, after which the background processes
become larger than the signal for  the cuts  considered. 
It has been checked that, in the channel $e^+ e^- \to 
\mu^+ \mu^- \tau^- \bar\nu_{\tau} u \bar d$, the high-energy behaviour of the cross section is in good
agreement with the one found in ref.~\cite{barger}.  For $\sqrt s = $~360 and 500~GeV also the combined effect
of ISR and beamstrahlung has been studied. For the full signal+background inclusive cross  sections, 
radiative effects turn out to be confined within a few per cent, being
dominated by ISR. This can be easily understood, since the Born approximation inclusive cross section is 
almost flat, as a function of $\sqrt s$, above, say, 300~GeV.

\begin{figure}[hbtp]
\begin{center}
\epsfig{file=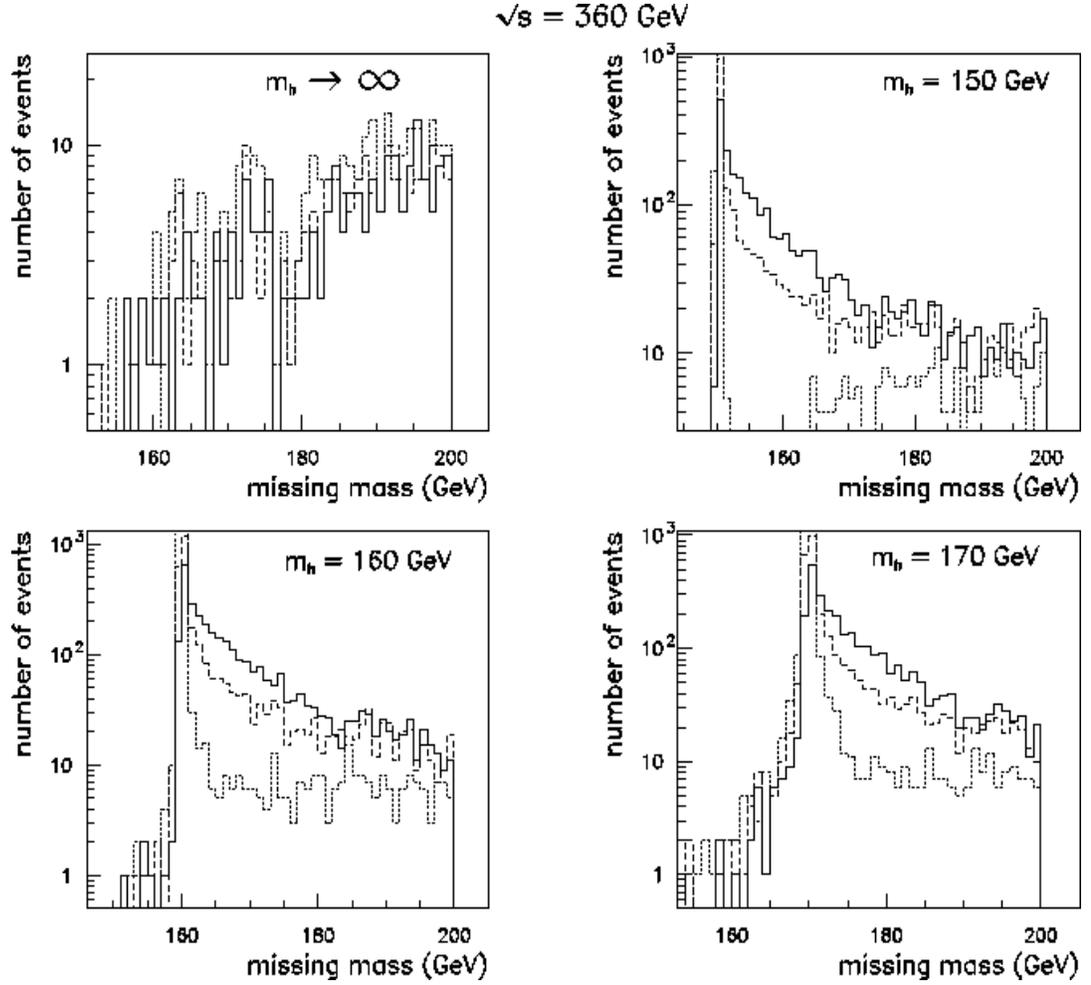,height=13truecm, width=13truecm}
\end{center}
\caption{The missing mass  distributions at $\protect{\sqrt{s}}  = 360$~GeV. Dotted, dashed and continuous
histograms correspond to Born approximation, Born+ISR and Born+ISR+beamstrahlung results, respectively.}
\label{fig:missmass360}
\end{figure}

\begin{figure}[hbtp]
\begin{center}
\epsfig{file=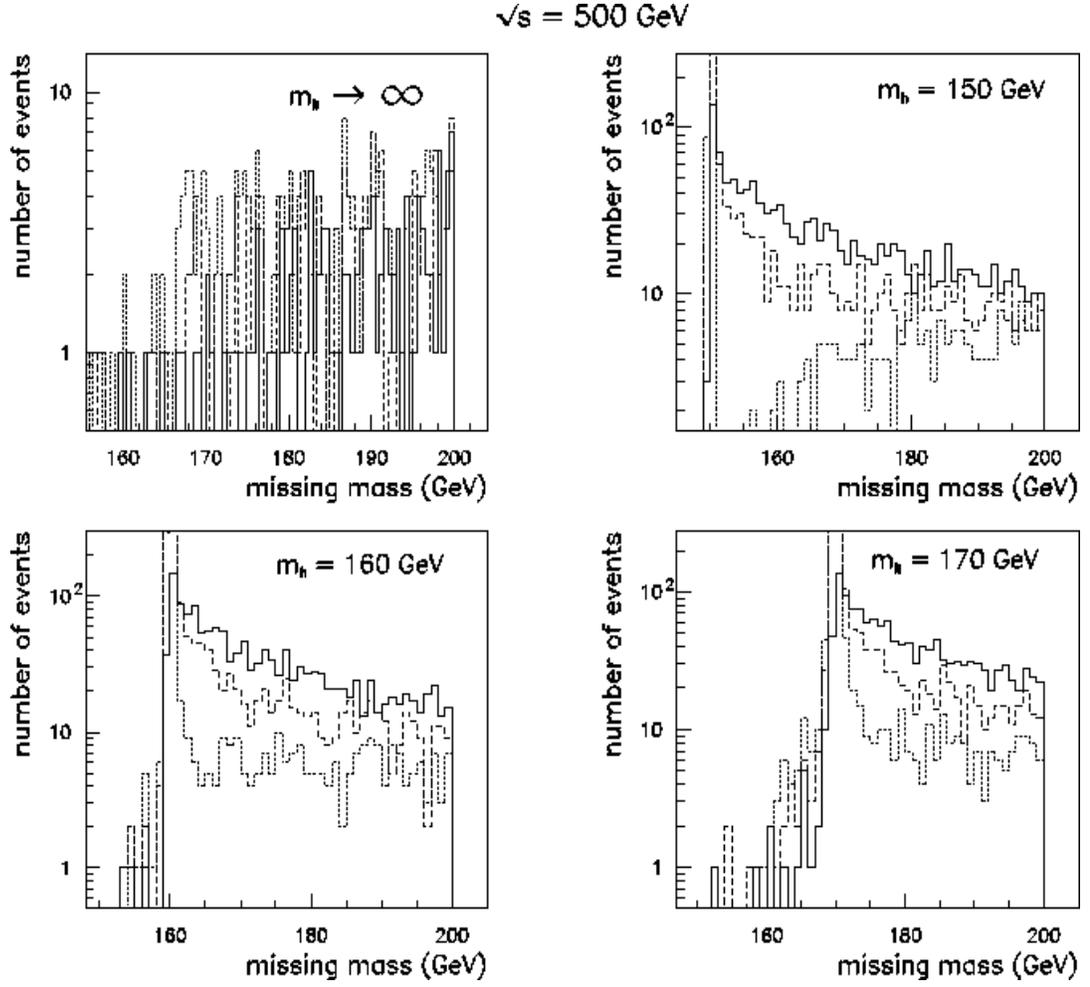,height=13truecm, width=13truecm}
\end{center}
\caption{ The missing mass  distributions at $\protect{\sqrt{s}}  = 500$~GeV. Notation as in
fig.~\ref{fig:missmass360}.}
\label{fig:missmass500}
\end{figure}

\begin{figure}[hbtp] 
\begin{center}
\epsfig{file=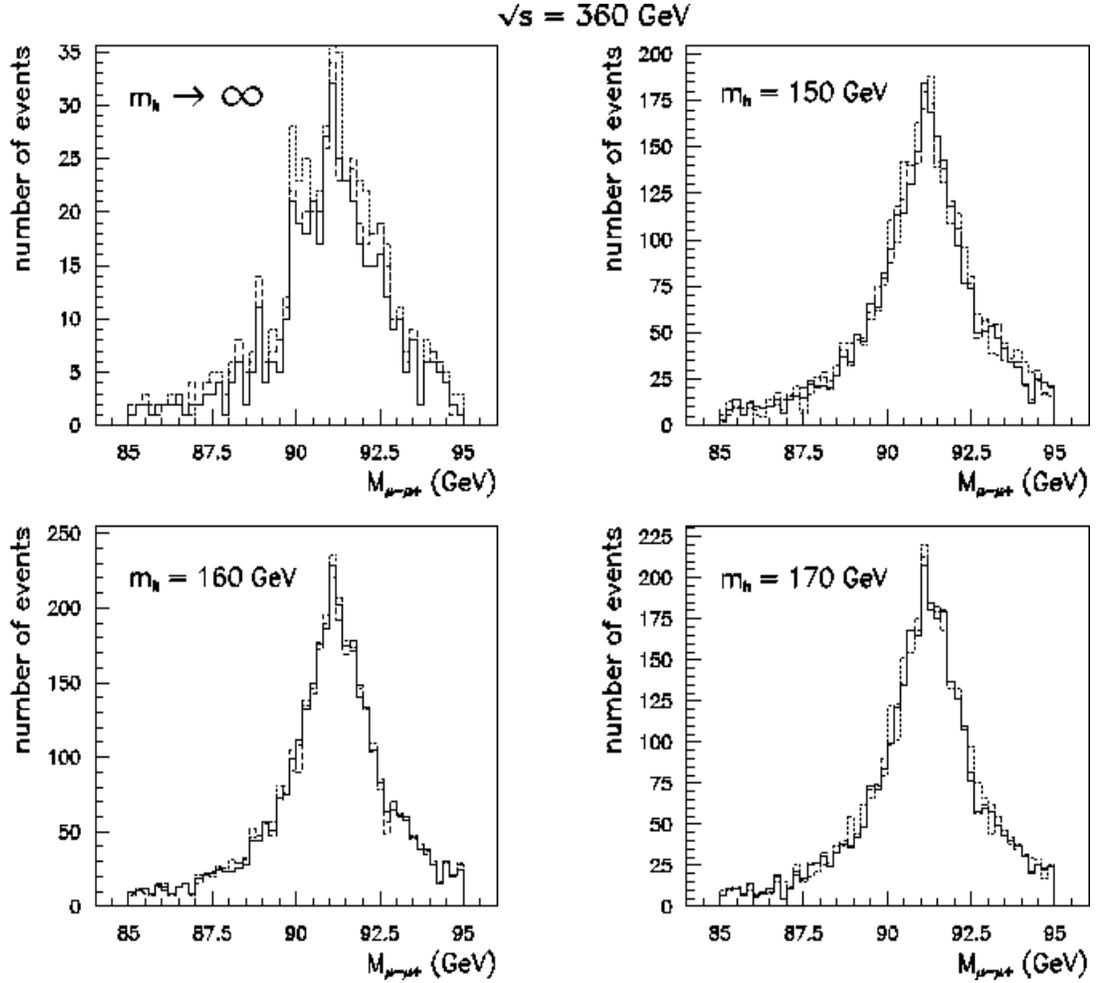,height=13truecm, width=13truecm}
\end{center}
\caption{The $\mu$-pair invariant mass distribution in the surroundings of $M(\mu^+ \mu^-) = M_Z$. 
Notation as in
fig.~\ref{fig:missmass360}.}
\label{fig:mumass}
\end{figure}

\begin{figure}[hbtp]
\begin{center}
\epsfig{file=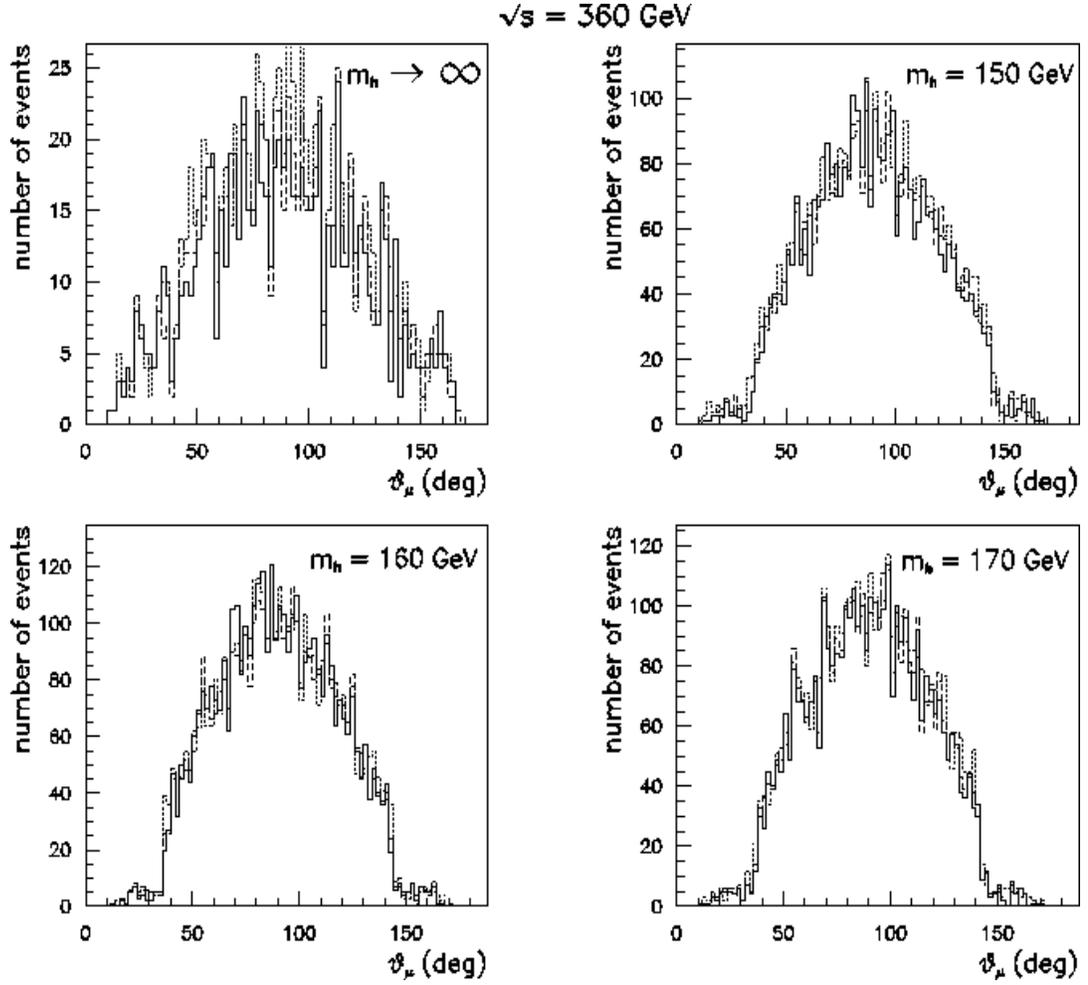,height=13truecm, width=13truecm}
\end{center}
\caption{The distribution of the angle  between the total $\mu$-pair  
three-momentum and the beams. Notation as in
fig.~\ref{fig:missmass360}.}
\label{fig:zangle}
\end{figure}

The results shown in figs.~\ref{fig:missmass360} - \ref{fig:correjet500} 
correspond to $\sqrt{s} = 360$~GeV, except for figs.~\ref{fig:missmass500} and
\ref{fig:correjet500}, where $\sqrt{s} = 500$~GeV. 
They have been obtained by analyzing the six-fermion unweighted event samples 
simulated by using the program as a generator. In all the cases, 
the number of events shown is normalized to a
fixed integrated luminosity,  
in order to perform a comparison consistent with the corresponding cross sections. 
The channel $e^+ e^- \to \mu^+ \mu^- e  \bar\nu_{e} u \bar d$  has 
been considered. As general comments, in these  figures the dotted histograms represent the Born approximation
results, the dashed ones represent the results including the effect of ISR and the solid ones represent the
full prediction, including also the effect of beamstrahlung.  
It has to be noticed that, among the various distributions shown, the
only ones that are very sensitive to the effect of ISR and/or beamstrahlung are the ones 
analyzed in figs.~\ref{fig:missmass360} and \ref{fig:missmass500}.

Figure~\ref{fig:missmass360} shows the distribution of the missing mass,  defined as $M_{miss}  = \sqrt{(P  - 
Q_z)^2}$, $P$ being the total incoming nominal four-momentum, i.e. before the energy losses due to ISR  and/or
beamstrahlung. In the absence of  radiative effects,  the missing mass coincides with the invariant mass of
the Higgs boson decay products. This is an experimentally  relevant quantity, since the Higgs boson width is
too tiny to be resolved directly. The radiative  effects amount generically to populate the high-mass tails of
the distribution. A comparison between figs.~\ref{fig:missmass360} and \ref{fig:missmass500} shows that such
effect is larger at $\sqrt{s} = 500$~GeV.

In fig.~\ref{fig:mumass} the invariant mass of the $\mu$ pair system is shown,  in a window of some $Z$ boson
widths around the $Z$ boson mass. The distribution is  always peaked at the  $Z$ boson mass; the presence of
the Higgs boson signal is revealed by the height of the peak, which is enhanced by a factor of about 6 with
respect to  the background.

Figure~\ref{fig:zangle} shows the distribution of  the angle between the total $\mu^+ \mu^-$ three-momentum
and the beam. In the presence of the Higgs boson signal only, it coincides with the $Z$ boson scattering
angle. As the Higgs boson mass varies, the distributions are quite similar, all smoothly peaked around
$\vartheta_\mu = \pi / 2$. The presence of the Higgs boson signal is revealed by the height of  the peak  also
in this case.

In fig.~\ref{fig:corre} the distribution of the variable $\xi_1 = (\cos \vartheta^*_{eu} + 
\cos\vartheta^*_{ed})  / 2$ is shown. The angles $\vartheta^*$ are defined as follows. First, the $\bar
\nu_e$ ``three-momentum'' is reconstructed as $\vec{v}_6 = - \sum_{i=1}^5 \vec{q}_i$ and its energy component is
defined by means of the shell relation $v_6^0 = \vert \vec{v}_6\vert $.  $v_6$ coincides
with the true $\bar \nu_e$ four-momentum only in the Born approximation;  in the presence of radiative 
effects in the collinear approximation $\vec{v}_6$ is the total lost three-momentum. Then the four vector $h$
defined as $h = q_3+q_4+q_5+v_6$ is considered; in the Born approximation and in the presence of the Higgs boson
signal only, it coincides with the Higgs boson four-momentum.  The angles $\vartheta^*$ are measured in the
reference frame in which $\vec{h} = \vec{0}$; in the Born approximation and in the presence of the Higgs boson
signal alone, it is the rest frame  of the Higgs particle. 
In the presence of the Higgs boson signal,  the distribution of $\xi_1$ sharpens around zero. 
In the above procedure,
there is an obvious arbitrariness in defining $v_6^0$; actually, it could also be defined by means of the
energy-momentum conservation, in such a way that $v_6$ is the total lost four momentum. It has been checked that
the results are not critical with respect to this choice. It has to be noticed that the variable $\xi_1$ is a
particularly sensitive variable; in fact, the presence of the Higgs boson signal is revealed by a significant
change in the shape of the distribution. 

Figures~\ref{fig:correjet360} and \ref{fig:correjet500} show the 
distributions of the variable $\xi_2$ defined
in an analogous way to $\xi_1$ as $\xi_2 = \vec{q}^*_5 \cdot \vec{Q}^*_{had}/q_5^{0*}/Q_{had}^{0*}$, where  
$q_5^*$ and $Q_{had}^*$ are the four-momenta of the outgoing electron and hadronic system, respectively, 
in the reference frame defined above. Incidentally, $Q_{had}$ coincides with the $W^+$ boson four-momentum. The
c.m. energies considered are $\sqrt{s} = 360$ and 500~GeV. For both the c.m. energies, the
variable $\xi_2$ shows an enhanced sensitivity to the presence of the Higgs boson particle.  Actually, the
background distribution has  a  more pronounced peak in the region $\xi_2 \simeq -1$. It has to be noticed that
at $\sqrt{s} = 500$~GeV there is a two-peak structure, where  the peak around $\xi_2 \simeq 0  $ is due to the
Higgs boson signal, whereas the one at $\xi_2 \simeq -1$ is due to  the background processes, which in this 
case give a contribution to the cross section comparable to the signal  contribution  (see
fig.~\ref{fig:totxsect}).  

The shapes of the variables $\xi_{1,2}$ are very sensitive to the presence of the Higgs boson signal. However, 
they are only partly determined by the underlying dynamics: in fact they are not sensitive to the spinless
nature of the Higgs particle. In order to point out the spinless nature of the Higgs boson, it is necessary to 
investigate more exclusive forms of angular correlations. To this aim, the variable $\xi_3$ has been defined as
$\cos \vartheta^*_{eu}$ or $\cos \vartheta^*_{ed}$, i.e. the cosine of the angle between the charged
lepton and the $up$- or $down$-quark in the reference frame defined above. Figures~\ref{fig:correud360} and 
\ref{fig:correud500} show the distribution of the two cosines for $\sqrt s = $~360~GeV and 500~GeV,
respectively. In all the histograms the effect of ISR and beamstrahlung have been included. 
The shaded and hatched histograms represent $\cos \vartheta^*_{eu}$ and $\cos \vartheta^*_{ed}$,
respectively. The white histogram represents the sum of the previous histograms. 
In  all the cases, $\cos \vartheta^*_{eu}$
is peaked at $-1$, both in the presence and in the absence of the Higgs boson signal. On the contrary, 
$\cos \vartheta^*_{ed}$ is peaked at $-1$ in the presence of the background processes only, whereas in the
presence of the Higgs boson signal the region $\cos \vartheta^*_{ed} \simeq 1$ is significantly populated. 
The shapes of the variables $\xi_3$ are sensitive to the spinless nature of the Higgs particle. 
 Such a property has been pointed out also in  a recent paper~\cite{dd}, as a
useful  tool in the search for the Higgs boson in the decay channel $h \to W^+ W^- \to l^+ l'^- \nu_l
\bar\nu_{l'}$ at the LHC. Anyway, in that paper the study has been performed on signal  and background
separately, without relying upon a full multi-particle calculation. It is worth noting that the sensitivity of
the shape is definitely larger at $\sqrt s = $~360~GeV rather than at $\sqrt s  = $~500~GeV, where signal and
background are  of the same order (see fig.~\ref{fig:totxsect}).  

\vskip 12pt\noindent
\section{Conclusions and outlook}

The production of a Standard Model Higgs boson 
in $e^+ e^-$ collisions at very high energies  has been studied, 
considering the case of intermediate-mass Higgs boson which yields 
 a six-fermion final state. The calculation of the Higgs boson signal
as well as of the background  matrix elements has been carried out
using the automatic algorithm {\tt ALPHA}, which reveals to be 
particularly powerful for such six-fermion calculations involving in the 
conventional diagrammatic approach a huge number of graphs. The tree-level 
scattering amplitudes returned by {\tt ALPHA} have been corrected for the 
effects of bremsstrahlung and beamstrahlung, in order to provide as 
comprehensive as possible results, by interfacing 
{\tt ALPHA} with the  Monte Carlo program {\tt HIGGSPV/WWGENPV}, properly adapted to treat 
$6f$ production processes.

\begin{figure}[hbtp]
\begin{center}
\epsfig{file=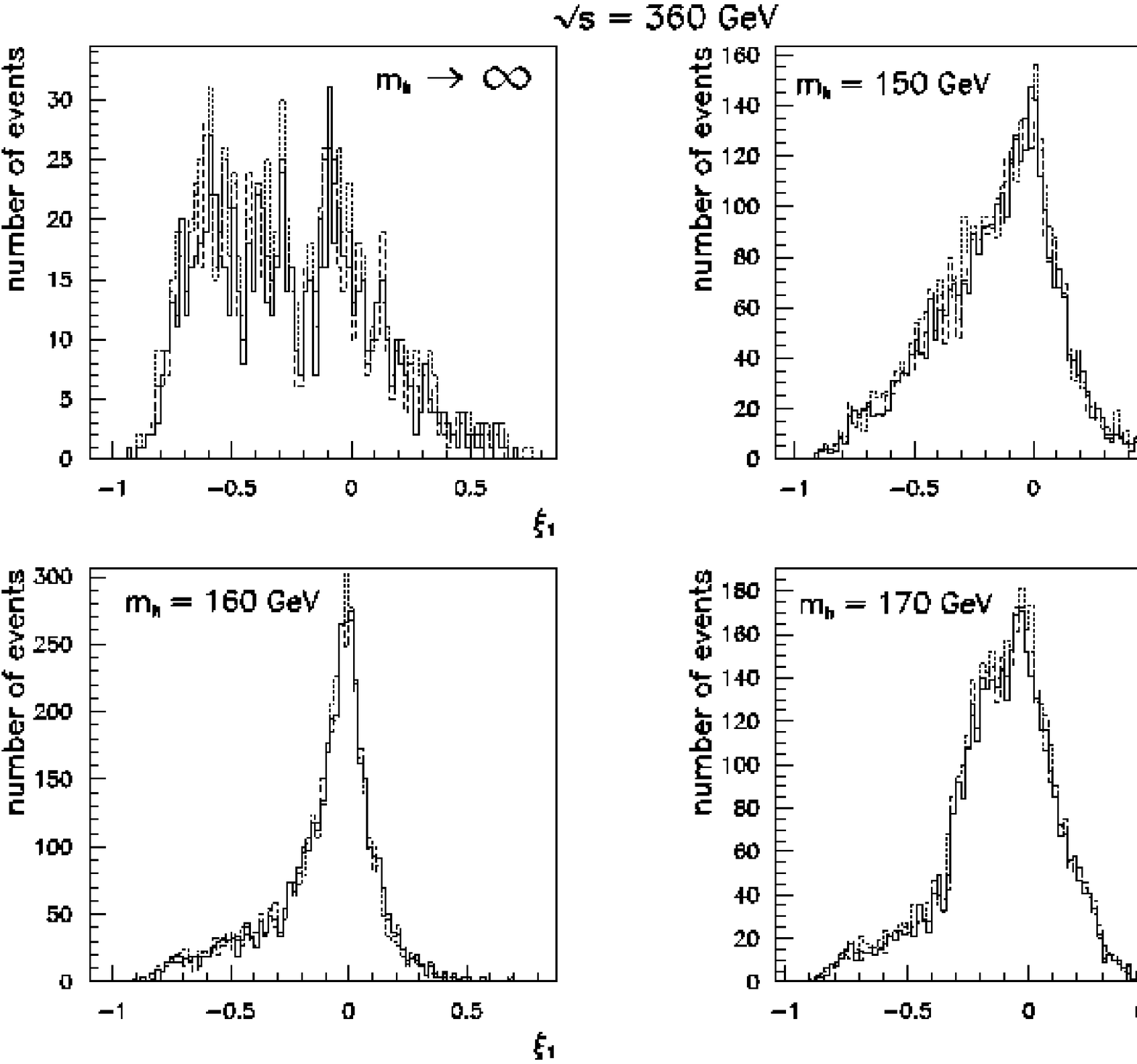,height=13truecm, width=13truecm}
\end{center}
\caption{The distribution of the variable $\xi_1$ (see the text for  the 
definition) at $\protect{\sqrt{s}} = 360$~GeV. Notation as in
fig.~\ref{fig:missmass360}.}
\label{fig:corre}
\end{figure}

\begin{figure}[hbtp]
\begin{center}
\epsfig{file=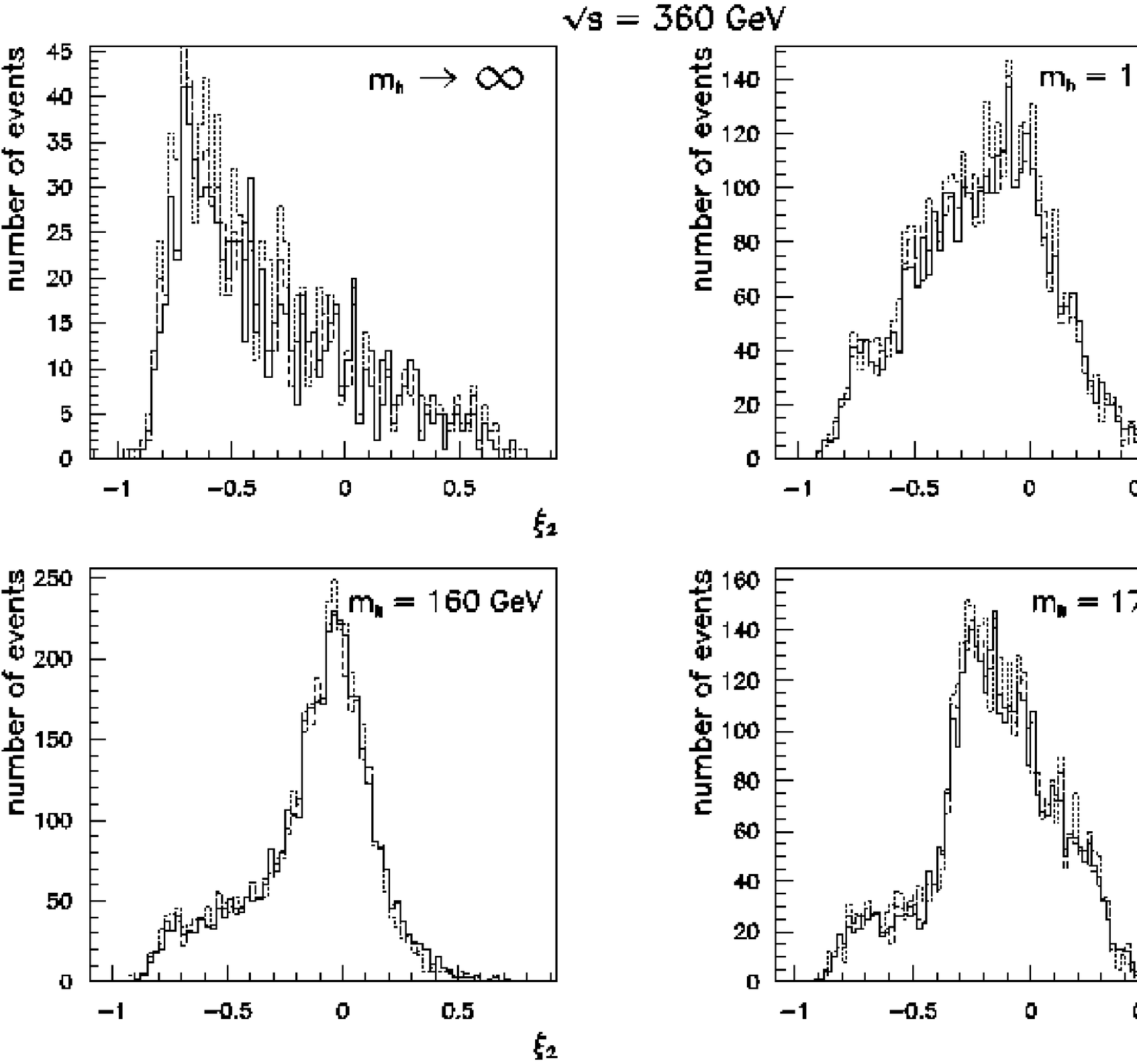,height=13truecm, width=13truecm}
\end{center}
\caption{The distribution of the variable $\xi_2$ (see the text for 
the definition) at $\protect{\sqrt{s}} = 360$~GeV. Notation as in
fig.~\ref{fig:missmass360}.}
\label{fig:correjet360} 
\end{figure}

\begin{figure}[hbtp]
\begin{center}
\epsfig{file=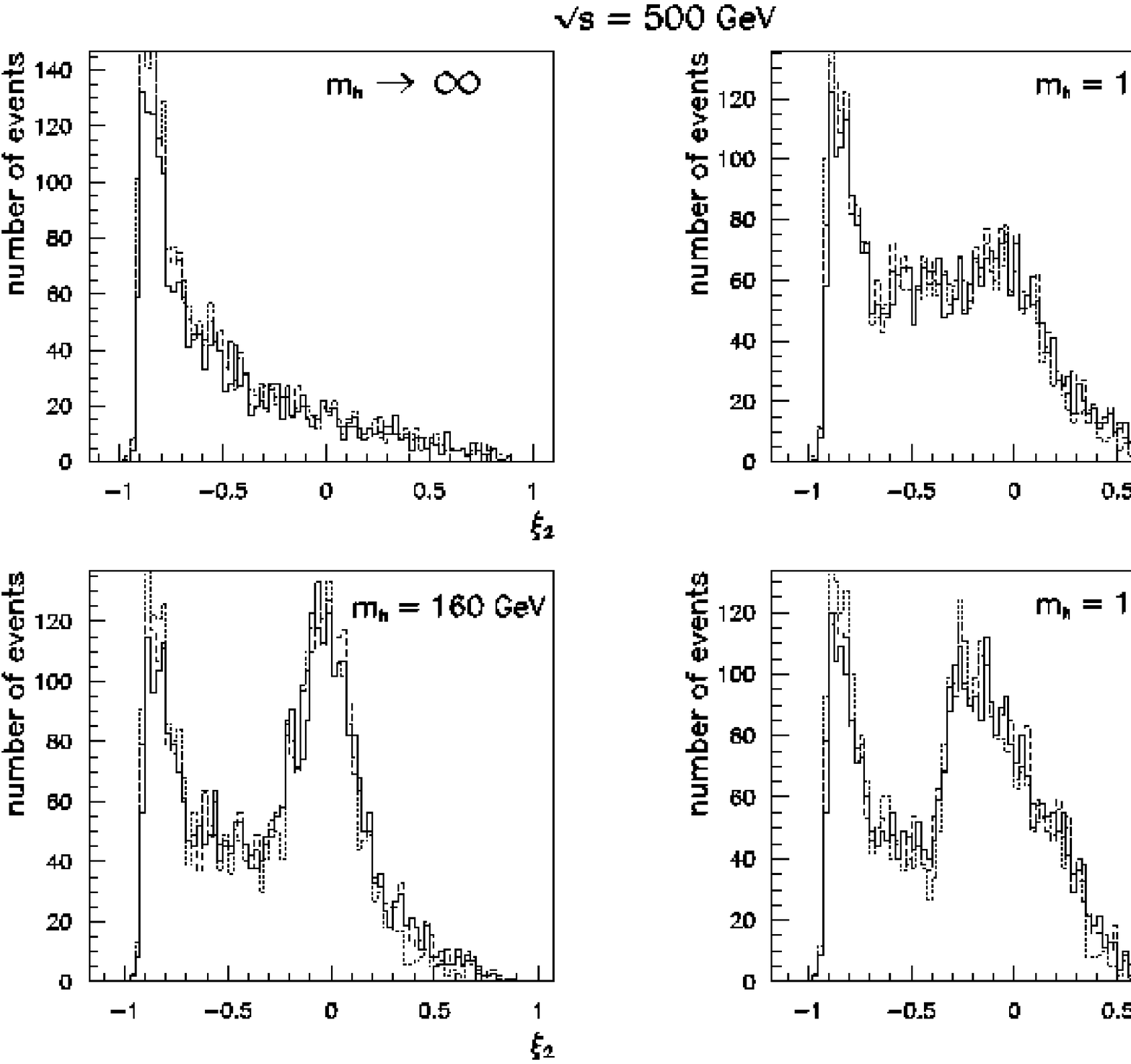,height=13truecm, width=13truecm}
\end{center}
\caption{The same as fig.~\ref{fig:correjet360} at $\protect{\sqrt{s}} = 
500$~GeV. Notation as in
fig.~\ref{fig:missmass360}.}
\label{fig:correjet500}
\end{figure}

\begin{figure}[hbtp]
\begin{center}
\epsfig{file=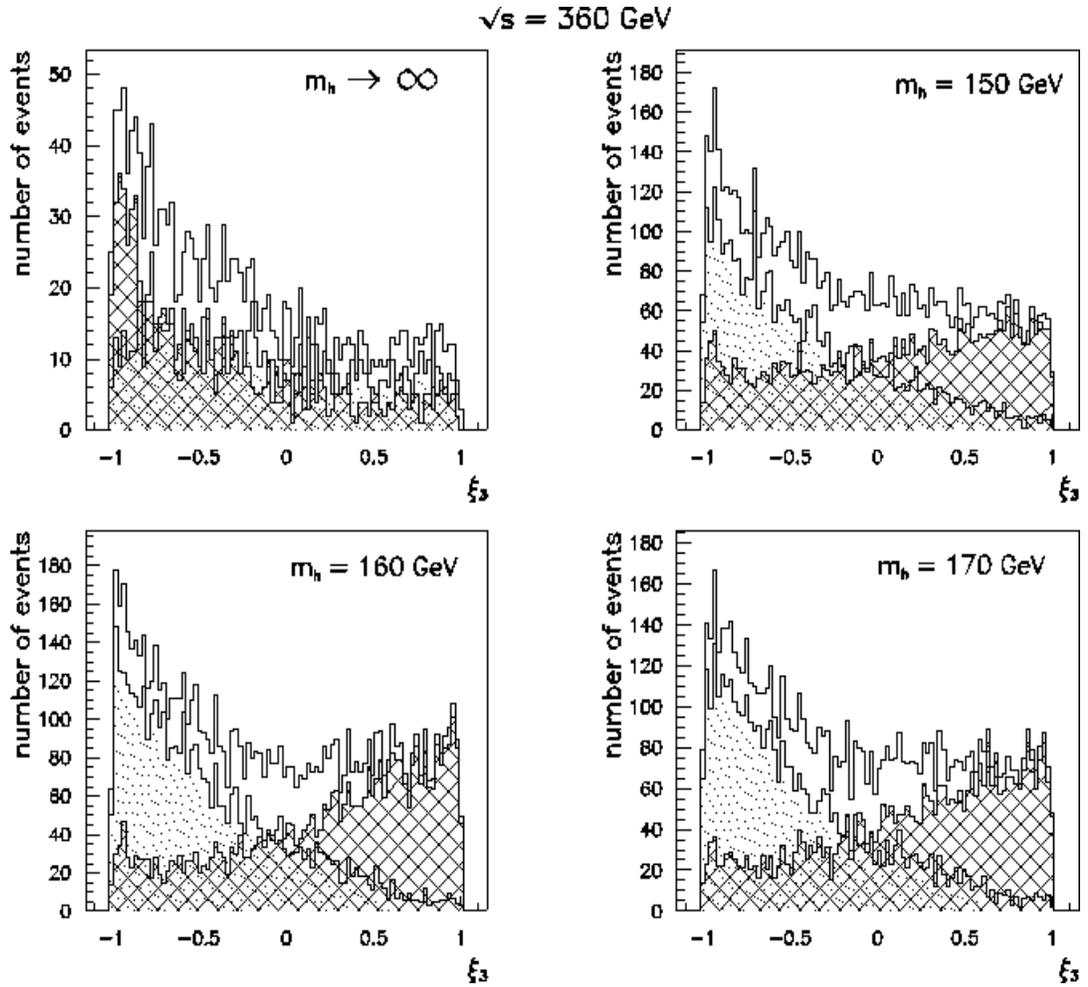,height=13truecm, width=13truecm}
\end{center}
\caption{The distribution of the variable $\xi_3$ (see the text for 
the definition) at $\protect{\sqrt{s}} = 360$~GeV.
All the histograms include the effect of ISR and beamstrahlung. The shaded and  hatched  
histograms  represent $\cos \vartheta^*_{eu}$  and $\cos \vartheta^*_{ed}$;  the white histogram 
is the sum of the previous ones. 
}
\label{fig:correud360}
\end{figure}

\begin{figure}[hbtp]
\begin{center}
\epsfig{file=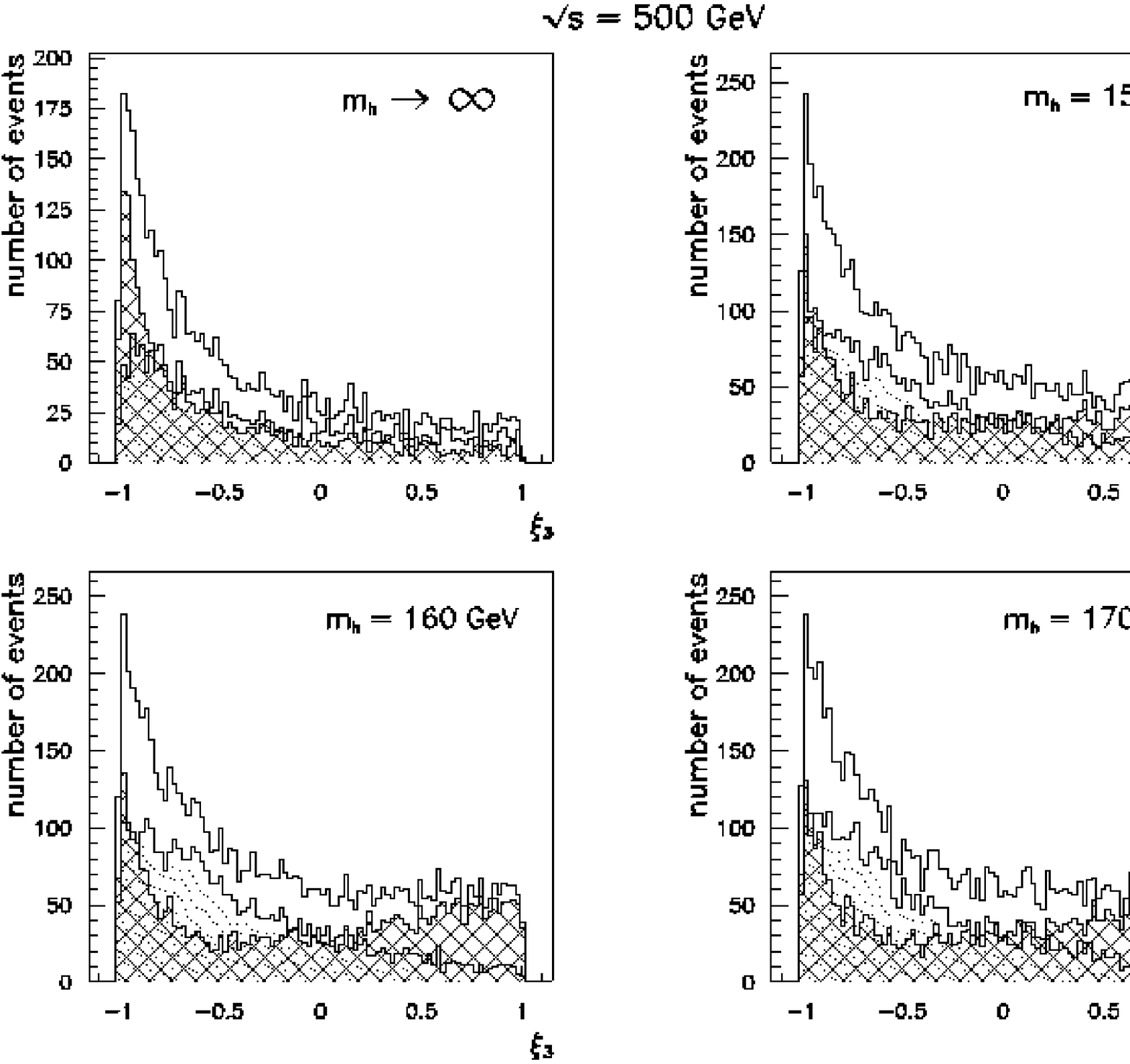,height=13truecm, width=13truecm}
\end{center}
\caption{The same as fig.~\ref{fig:correud360} at $\protect{\sqrt{s}} = 
500$~GeV. Notation as in fig.~\ref{fig:correud360}. }
\label{fig:correud500}
\end{figure}

As a summary of  the main outcomes of the study, the following items  have to  be pointed out:
 
\begin{itemize}

\item the full $6f$ calculation as compared with the calculation carried out within the 
narrow-width approximation shows that the various off-shellness effects can be of the order of
some per cent; for experimental accuracies better that some per cent  the narrow-width
approximation becomes inadequate; 

\item the full calculation as compared to the incoherent sum of signal and background reveals
that, for the channels considered, signal/background interference effects  are compatible with 
zero at the level of precision of about 0.5\%;   

\item the missing mass distribution, relevant for the determination of the Higgs  boson mass,
is  strongly affected by initial-state radiation and/or beamstrahlung effects (radiative
effects), as expected; 

\item the shapes of all the  other distributions considered  in the paper are only smoothly
affected by radiative effects; 

\item the variables that are mostly sensitive to the presence of the Higgs particle decaying  into  a
$W$-boson  pair are $\xi_1$ and $\xi_2$ of figs.~\ref{fig:corre}, \ref{fig:correjet360} and
\ref{fig:correjet500}; the  variable most  sensitive to the spinless nature of the Higgs boson is  $\xi_3$
of figs.~\ref{fig:correud360} and \ref{fig:correud500}; in all the cases, they 
are variables concerning the spin correlations of  the Higgs boson
decay products, and can  be meaningfully analyzed only by means of a complete $6f$ calculation.  

\end{itemize}

As a conclusion, the approach described in the present paper represents a valuable tool for
Higgs boson searches at NLC and, more generally, for 
the study of scattering processes involving multi-particle production.  

\vskip 12pt \noindent
{\bf Acknowledgements} One of  the authors (M.M.) thanks the 
Dipartimento di Fisica Nucleare e
Teorica, University of Pavia, for financial support and hospitality during the development of
the present work, and the INFN, Sezione di Pavia,
for allowing the use of computing facilities.

\vfil \eject

\bibliography{sixf}

\newpage

\newpage

\newpage

\newpage

\newpage

\newpage

\newpage

\newpage

\newpage

\newpage

\newpage

\newpage

\newpage

\newpage

\newpage

\end{document}